\documentclass[a4paper,11pt]{article}
\pdfoutput=1 %\documentclass[a4paper,11pt]{article}
\usepackage{amssymb,mathrsfs}
\usepackage{bbm,bm}
\usepackage{siunitx}
\usepackage{amsmath}
\usepackage{braket}
\usepackage{mathtools}
\usepackage[usenames,dvipsnames,svgnames,table]{xcolor}
\usepackage [utf8] {inputenc}
\usepackage{color}
\usepackage{lmodern}
\usepackage{footnote}
\usepackage{slashed}
 \usepackage{mathrsfs}
 \usepackage[pdftex] {graphicx}
\usepackage{multirow}
\usepackage{jheppub} 
\usepackage{here}
\usepackage{hyperref}
\usepackage[justification=centering, singlelinecheck=false]{caption}
\usepackage[list=true, labelfont=bf, labelformat=brace, position=top]{subcaption}
\newcommand{\eq}{\begin{eqnarray}}
\newcommand{\en}{\end{eqnarray}}

\title{Modified L\"uscher zeta-function and the modified effective range expansion in the presence of
a long-range force}

\renewcommand{\thefootnote}{\fnsymbol{footnote}}

\author[1]{Rishabh Bubna,}
\affiliation[1]{Helmholtz-Institut f\"ur Strahlen- und Kernphysik (Theorie)\\ and Bethe Center for Theoretical Physics, Universit\"at Bonn, 53115 Bonn, Germany}
\emailAdd{bubna@hiskp.uni-bonn.de}

\author[2,3]{Hans-Werner Hammer,}

\affiliation[2]{Technische Universit\"at Darmstadt, Department of Physics,
64289 Darmstadt, Germany}
\affiliation[3]{ExtreMe Matter Institute EMMI and Helmholtz Forschungsakademie
  Hessen f\"ur FAIR (HFHF),
GSI Helmholtzzentrum f\"ur Schwerionenforschung GmbH,
64291 Darmstadt, Germany}
\emailAdd{hans-werner.hammer@physik.tu-darmstadt.de}

\author[4]{Bai-Long Hoid,}
\affiliation[4]{
Institut f\"ur Kernphysik and PRISMA$^+$  Cluster of Excellence, Johannes Gutenberg Universit\"at,  55099 Mainz, Germany}
\emailAdd{lonbai@uni-mainz.de}

\author[5]{Jin-Yi Pang,\footnote{Corresponding author.}}
\affiliation[5]{College of Science, University of Shanghai for Science and Technology, Shanghai 200093, China}
\emailAdd{jypang@usst.edu.cn}

\author[1,6]{Akaki Rusetsky,}
\affiliation[6]{Tbilisi State  University,  0186 Tbilisi, Georgia}
\emailAdd{rusetsky@hiskp.uni-bonn.de}

\author[7]{and Jia-Jun Wu}
\affiliation[7]{School of Physical Sciences, University of Chinese Academy of Sciences, Beijing 100049, China}
\emailAdd{wujiajun@ucas.ac.cn}

\abstract{

\noindent
An efficient numerical algoritm is proposed for the calculation of the modified L\"uscher
zeta-function in the presence of a long-range force. Using the formalism developed in
Ref.~\cite{Bubna:2024izx} for the analysis of synthetic data on the finite-volume energy
levels in a toy model, it is demonstrated that, in contrast to the standard L\"uscher 
approach, the truncation of the higher partial waves has very little effect on the final
result.  Furthermore, the regularization and renormalization of the modified L\"uscher
zeta-function is discussed in detail, as well as the problems arising within the cutoff
regularization. It is shown that, using the renormalization scheme proposed in the present
paper, one obtains modified effective range expansion parameters of natural size
in all partial waves.
}

\allowdisplaybreaks
\begin{document}
\maketitle
\flushbottom

\renewcommand{\thefootnote}{\arabic{footnote}}

\section{Introduction}

The recent surge of interest to the two-body finite-volume quantization condition
in the presence of a long-range
force~\cite{Meng:2021uhz,Meng:2023bmz,Raposo:2023oru,Bubna:2024izx,Raposo:2025dkb,Hansen:2024ffk,Dawid:2024dgy,Dawid:2024oey,Abolnikov:2024key,Du:2025vkm,Dawid:2025wsn,Yu:2025gzg}
can be attributed, first and foremost, to the ongoing 
studies of two specific problems in lattice QCD. These are the calculation of the
$D^*D$ spectrum and the properties of the $T_{cc}^+(3875)$
state~\cite{Whyte:2024ihh,Collins:2024sfi,Prelovsek:2025vbr,Meng:2024kkp}, as well as to the extraction
of the baryon-baryon scattering phase shifts from the two-baryon
spectrum~\cite{Green:2021qol,Amarasinghe:2021lqa,Green:2025rel}.\footnote{For an alternative approach to the study of these systems by using HAL QCD method, see, e.g.,~\cite{Aoki:2020bew,Lyu:2023xro,Aoki:2025jvi}.}
At physical quark masses, both these systems feature a long-range force that emerges from
the one-pion exchange. This leads to several interrelated problems in the analysis of lattice
data, all stemming from the fact that the scale that defines the effective range of
interactions could become sizable as compared to the inverse of the box size $L$.
(For the systems in question, this scale is of order of the pion mass or even less.)
Namely, the exponentially suppressed corrections, which are usually neglected in the
derivation of the L\"uscher equation~\cite{Luscher:1990ux}, may become
relevant. Furthermore, the standard L\"uscher approach relies on the truncation of higher
partial waves. However, as one knows, the convergence of the partial-wave expansion
in the presence of a long-range force is poor, and many terms have to be kept in this
expansion in order to achieve a reasonable accuracy. Last but not least, it is well known
that the partial-wave scattering amplitudes in quantum field theory develop the
so-called  left-hand cut in the complex $s$-plane. In case of $NN$ scattering, the beginning of this cut lies just 5\,MeV below the elastic $NN$ threshold.
In general, the width of the gap between left- and right-hand cuts is related to the effective
range of the interactions and is very small for the systems considered here.
It is also clear that, for the analysis of the energy levels which lie on or in the vicinity of the
left-hand cut, the standard L\"uscher method is not applicable.

In recent years, several approaches have been proposed to address the problems
mentioned above. The work done within the HAL QCD
approach~\cite{Aoki:2020bew,Lyu:2023xro,Aoki:2025jvi}
is conceptually
very different from the rest and will not be considered here. Furthermore, a very straightforward approach, uses an effective field theory based Hamiltonian and treats long-
and short-range interactions on equal footing ~\cite{Meng:2021uhz}. This approach
parameterizes the interactions in terms of effective couplings which are, by definition,
devoid of any singularities, in contrast to the $S$-matrix elements. The scattering
equations are then solved directly in a finite volume, producing the finite-volume spectrum
that is fitted to the data. In order to avoid large partial-wave mixing effects and the
emergence of the left-hand cut, the equations are solved in the plane-wave basis, without
resorting to the partial-wave expansion. In this way, one also takes into account exponentially
suppressed corrections, apart from those that are implicitly included in the
effective couplings and are determined by a heavy scale. At the final stage, one solves
the same equation in the infinite volume in order to determine $S$-matrix elements,
using the values of the couplings extracted from the fit to the lattice spectrum.
We further note that the approach, where the two-body scattering is
``embedded'' into the three-body problem
(see, e.g.,~\cite{Hansen:2024ffk,Dawid:2024dgy}), is based on similar premises. Namely,
below the breakup threshold, it is equivalent to the two-body equation written down
in the plane-wave basis, where the sum is performed over the spectator three-momentum,
while, of course, these two approaches differ above the breakup threshold.
An alternative approach, using the finite-volume version of
the $N/D$ method, has also been proposed recently~\cite{Dawid:2024oey}.

Another group of approaches is based on the explicit splitting of the interaction into
the known long-range and the unknown short-range
parts~\cite{Raposo:2023oru,Bubna:2024izx,Raposo:2025dkb}. In particular, in this paper
we shall concentrate on the approach developed in Ref.~\cite{Bubna:2024izx}, which
uses the finite-volume generalization of the so-called modified effective range
expansion by van Haeringen and Kok~\cite{vanHaeringen:1981pb}. Here, the long-range
interaction is treated in the plane-wave basis, whereas the partial-wave expansion
for the short-range part is still performed. This expansion is assumed to be converging well,
which will be explicitly verified below. As a result, one obtains a modified L\"uscher
equation with a little partial-wave mixing, which allows one to extract the scattering phase
pretty much in the same way as in the standard L\"uscher approach. We would like
to stress that we do not plan a comparison of different approaches with the aim
of determining the best one. As far as we can see, all of them are conceptually equivalent
and the decision of which one to use should be based on mere convenience, see Ref.~\cite{Dawid:2025wsn} for more discussion on this issue.

The aim of the present paper is to demonstrate the numerical implementation of the
formalism developed in Ref.~\cite{Bubna:2024izx}, carrying out the analysis of synthetic
data generated in a toy model. In the course of this, we shall address all issues mentioned
in the beginning, namely, the size of the exponentially suppressed corrections,
partial-wave mixing, and the analysis of the data in the region of the left-hand cut.
Another important issue that we consider in the present article is the renormalization.
In particular, the modified L\"uscher zeta-function, which enters the modified L\"uscher
equation, is given by an infinite sum of diagrams corresponding to the multiple insertions
of the long-range potential into a two-particle loop. These diagrams are
ultraviolet-divergent, with the degree of divergence growing with the angular momentum
of a given partial wave. It can be seen that the cutoff regularization becomes inconvenient
in higher partial waves, since power-law divergences force the parameters of the
effective range expansion to be of an unnatural size. We therefore opt for the dimensional
regularization and provide an efficient numerical algorithm for the calculation of the
above-mentioned multi-loop integral in this regularization. It should be stressed that
the results of these studies are relevant beyond the applications in the analysis of
the finite-volume lattice data and can be used for the study of the few-body systems
in the presence of several interactions with very different scales.

The layout of the paper is as follows. In order to render the paper self-contained, we review the key formulae from Ref.~\cite{Bubna:2024izx} in Sect.~\ref{sec:compendium}. This material is essential for the understanding of the rest of the paper. In Sect.~\ref{sec:luescher}, we consider the calculation of the modified
L\"uscher zeta-function and the analysis of the (synthetic) data. Finally, in
Sect.~\ref{sec:dimensional}, we consider the renormalization of the modified L\"uscher
zeta-function in dimensional regularization and the modified effective range expansion.
To separate the conceptual issues from unnecessary technical details, we work
exclusively in the center-of-mass frame and concentrate on the $A_1^+$ representation
of the octahedral group. Furthermore, we assume that the long-range part of the potential
is given by a simple Yukawa interaction corresponding to the exchange of a light particle
with a mass $M$. Going beyond these assumptions forms the subject of a separate
investigation.

\section{Modified effective range expansion and modified L\"uscher equation}
\label{sec:compendium}

In the infinite volume, the two-body scattering amplitude $T$ obeys the Lippmann-Schwinger
equation
\eq
T(\bm{p},\bm{q};q_0^2)=V(\bm{p},\bm{q})+\int\frac{d^3\bm{k}}{(2\pi)^3}\,
\frac{V(\bm{p},\bm{k})T(\bm{k},\bm{q};q_0^2)}{\bm{k}^2-q_0^2}\, ,
\en
where the variable $q_0^2$ is understood to have a small positive imaginary part, $q_0^2+i\epsilon$, above threshold as usual ($q_0$ denotes the magnitude of the relative three-momenta on shell). The potential $V=V_L+V_S$ is given by a sum of a long- and short-range parts.  
The long-range part is known and local. In the following, we assume that it is given by the Yukawa
potential corresponding to the exchange of a particle with the mass $M$ (the pion)
\eq\label{eq:longrange}
V_L(\bm{p},\bm{q})=\frac{4\pi g}{M^2+(\bm{p}-\bm{q})^2}\, .
\en
In case of the $NN$ scattering within the chiral effective theory, for example, in
the $^3P_0$ partial wave, $g=mM^2g_A^2/(16\pi F_\pi^2)\simeq 0.073\,m$, where
$m$, $g_A\simeq 1.27$ and $F_\pi\simeq 92.3\,\mbox{MeV}$
denote the nucleon mass, axial-vector coupling constant and the pion decay constant,
respectively. We shall use this value of $g$ in the toy model considered below.

The short-range part is generally unknown and non-local. Its expansion in momenta
takes the form
\eq
V_S(\bm{p},\bm{q})=C_0^{00}+3C_1^{00}\bm{p}\bm{q}+C_0^{10}(\bm{p}^2+\bm{q}^2)+\cdots\, .
\en

Note that in the derivation of the modified effective range expansion, it is usually assumed
that the long-range potential is superregular. In the partial wave with angular momentum
$\ell$ this means that $r^{-2\ell}V_L(r)$ stays analytic at $r\to 0$. The potential in
Eq.~(\ref{eq:longrange}) does not fulfill this condition, even for $\ell=0$. One could use
different types of regularizations here, e.g., Pauli-Villars regularization as
in~ Ref.\cite{Bubna:2024izx}, or merely a sharp cutoff. As we shall see, however, this method
is numerically inconvenient, when higher partial waves are included. The loops with
one-pion exchanges are divergent and produce high powers of the cutoff mass.
Renormalization implies that these large polynomial contributions are included
in the short-range couplings $C_0^{00}.\,C_1^{00},\,C_0^{10}\ldots$, rendering them
to be unnaturally large. We shall therefore argue in favor of the dimensional 
regularization and introduce a renormalization scheme where such large contributions
are absent from the beginning.

\begin{figure}[t]
  \begin{center}
    \includegraphics*[width=10.cm]{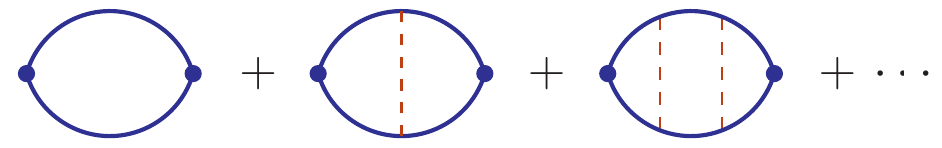}
  \end{center}
  \caption{Perturbative expansion of the function $M_\ell(q_0)$ defined in Eq.~(\ref{eq:M_ell}) in the coupling constant $g$. One-pion-exchange ladders to all orders contribute to this expansion.
    The filled dots correspond to the insertion of $\mathscr{Y}^*_{\ell m}(\bm{p})$
 and   $\mathscr{Y}_{\ell' m'}(\bm{q})$.}
  \label{fig:loops}
\end{figure}

The modified effective range expansion for the scattering phase shift on the full potential
$V$ can be written in the following form
\eq\label{eq:modified}
K_\ell^M(q_0^2)=M_\ell(q_0)+\frac{q_0^{2\ell+1}}{|f_\ell(q_0)|^2}\,
(\cot(\delta_\ell(q_0)-\sigma_\ell(q_0))-i)\, .
\en
Here, $K_\ell^M(q_0^2)$ denotes the so-called modified effective range function, whereas
$M_\ell(q_0)$ is given by
\eq\label{eq:M_ell}
\delta_{\ell\ell'}\delta_{mm'}
M_\ell(q_0)
=(4\pi)^2\int\frac{d^3\bm{p}}{(2\pi)^3}\,\frac{d^3\bm{q}}{(2\pi)^3}\,
\mathscr{Y}^*_{\ell m}(\bm{p})\langle\bm{p}|G_L(q_0)|\bm{q}\rangle
\mathscr{Y}_{\ell'm'}(\bm{q})\, ,
\en
where $\mathscr{Y}_{\ell m}(\bm{p})=p^\ell Y_{\ell m}(\hat p)$ and $Y_{\ell m}(\hat p)$ denotes the
conventional spherical function. Furthermore, $G_L(q_0)$ is the Green function for
the scattering on the long-range potential only, which obeys the following equation
\eq\label{eq:G_L}
\langle\bm{p}|G_L(q_0)|\bm{q}\rangle=\frac{(2\pi)^3\delta^{3}(\bm{p}-\bm{q})}
{\bm{p}^2-q_0^2}+\frac{1}{\bm{p}^2-q_0^2}\,
\int\frac{d^3\bm{k}}{(2\pi)^3}\,V_L(\bm{p},\bm{k})
\langle\bm{k}|G_L(q_0)|\bm{q}\rangle\, .
\en
Expanding the quantity $G_L(q_0)$ in perturbative series, it is seen that
$M_\ell(q_0)$ is given by the sum of loops with zero, one, two,\ldots insertions
of the one-pion exchange, see Fig.~\ref{fig:loops}. Of course, the above expressions do
not make sense as they stand, since the integrals entering some of the expressions diverge
in the ultraviolet. Some regularization/renormalization is implicit in these expressions.
Other quantities that enter into Eq.~(\ref{eq:modified}), are the phase shift
on the full potential $\delta_\ell(q_0)$, the phase shift on the long-range potential
$\sigma_\ell(q_0)$, and the Jost function for the long-range potential $f_\ell(q_0)$.
Thus, all quantities on the right-hand side of this equation, except $\delta_\ell(q_0)$,
can be evaluated from the known potential $V_L$ prior to performing any lattice
calculations. The rationale for splitting the potential into the long- and short-range parts
is manifested in the properties of $K_\ell^M(q_0^2)$. Namely, while the effective range
expansion of the conventional
$K$-matrix, obtained without such a splitting, has a very small radius of convergence due to the proximity of the left-hand cut, $K_\ell^M(q_0^2)$ stays regular
in the left-hand region.

The modified L\"uscher equation, which is based on the same splitting, performed in
a finite volume, is given by $\det\left[\mathscr{A}(q_0)\right]=0$, where
\eq\label{eq:A}
\mathscr{A}_{\ell m,\ell'm'}(q_0)=\frac{1}{4\pi}\,\delta_{ll'}\delta_{mm'}K_\ell^M(q_0^2)
-H_{\ell m,\ell'm'}(q_0)\, .
\en
In Eq.~(\ref{eq:A}), the quantity $H_{\ell m,\ell'm'}(q_0)$ is defined as (cf. with Eq.~(\ref{eq:M_ell}))
\eq
H_{\ell m,\ell'm'}(q_0)=\frac{4\pi}{L^6}\,\sum_{\bm{p},\bm{q}}
\mathscr{Y}^*_{\ell m}(\bm{p})\langle\bm{p}|G_L(q_0)|\bm{q}\rangle
\mathscr{Y}_{\ell'm'}(\bm{q})\, ,
\en
where $G_L(q_0)$ is given by Eq.~(\ref{eq:G_L}), with the integration replaced by
summation over the discrete momenta $\bm{p}=2\pi\bm{n}/L\,,\; \bm{q}=2\pi\bm{n'}/L\,,$ with $\bm{n}\,, \bm{n'}\in \mathbb{Z}^3$.

Owing to the lack of the rotational invariance, the finite-volume quantization condition
is not diagonal in angular momentum. Partial diagonalization is achieved using the
octahedral symmetry of the cubic lattice. To this end, one defines
(see, e.g.,~\cite{Luscher:1990ux,Bernard:2008ax,Gockeler:2012yj})
\eq\label{eq:A-Gamma}
\delta_{\sigma\rho}\mathscr{A}^\Gamma_{t\ell,t'\ell'}(q_0)
=\sum_{mm'}\left(c^{\Gamma\sigma t}_{\ell m}\right)^*
\mathscr{A}_{\ell m,\ell'm'}(q_0)c^{\Gamma\rho t'}_{\ell' m'}\, .
\en
Here, $\Gamma$ denotes a particular irreducible representation (irrep) of the octahedral
group, and $t,t'$ are indices that label different copies of the same irrep $\Gamma$
contained in the irreps of the rotation group characterized by angular momenta $\ell,\ell'$
respectively. Furthermore, $\sigma,\rho$ label basis vectors in the irrep $\Gamma$,
and $c^{\Gamma \sigma t}_{\ell m},\,c^{\Gamma\rho t'}_{\ell' m'}$ denote the pertinent Clebsh-Gordan
coefficients. The quantization condition in a given irrep is now given by
$\det\left[\mathscr{A}^\Gamma(q_0)\right]=0$. Finally, the spherical harmonics after the projection onto
the irrep $\Gamma$ are defined as
\eq
\mathscr{Y}^{\Gamma \sigma t}_\ell(\bm{p})=\sum_mc^{\Gamma\sigma t}_{\ell m}
\mathscr{Y}_{\ell m}(\bm{p})\, .
\en
Owing to the Wigner-Eckart theorem, the sum in the r.h.s. of Eq.~(\ref{eq:A-Gamma}) yields the factor $\delta_{\sigma\rho}$ with the coefficient that does not depend on $\sigma$ and $\rho$. For this reason, one may fix $\sigma=\rho$ to any value in the calculations. For this reason, and in order to avoid clutter of indices, in the following we suppress $\sigma,\rho$ in
all expressions.

\section{Modified L\"uscher zeta-function and the analysis of data}
\label{sec:luescher}

\subsection{Calculation of the zeta-function}

In this section we shall concentrate on the calculation of the quantity that enters
the matrix $\mathscr{A}^\Gamma_{t\ell,t'\ell'}(q_0)$ from Eq.~(\ref{eq:A-Gamma}):
\eq\label{eq:modified_qc}
H^\Gamma_{t\ell,t'\ell'}(q_0)=\frac{4\pi}{L^6}\sum_{{\bm p},\bm{q}}
\left(\mathscr{Y}^{\Gamma t}_\ell(\bm{p})\right)^*
\langle\bm{p}|G_L(q_0)|\bm{q}\rangle
\mathscr{Y}^{\Gamma t'}_{\ell'}(\bm{q})\, .
\en
In the infinite volume, the summation is replaced by integration, and we have
\eq\label{eq:HM}
H^{\Gamma,\infty}_{t\ell,t'\ell'}(q_0)
&=&\sum_{mm'}\left(c^{\Gamma t}_{\ell m}\right)^*
H_{\ell m,\ell' m'}^\infty(q_0)
c^{\Gamma t'}_{\ell' m'}
\nonumber\\[2mm]
&=&\frac{1}{4\pi}\,\sum_{mm'}\left(c^{\Gamma t}_{\ell m}\right)^*
\delta_{\ell\ell'}\delta_{mm'}
\mbox{Re}\left[M_\ell(q_0)\right]c^{\Gamma t'}_{\ell' m'}
\nonumber\\[2mm]
&=&\frac{1}{4\pi}\,\delta_{\ell\ell'}\delta_{mm'}\delta_{tt'}
\mbox{Re}\left[M_\ell(q_0)\right]\, .
\en
One further defines
\eq
\Delta H^\Gamma_{t\ell,t'\ell'}(q_0)
&=&H^\Gamma_{t\ell,t'\ell'}(q_0)-H^{\Gamma,\infty}_{t\ell,t'\ell'}(q_0)
\nonumber\\[2mm]
&=&H^\Gamma_{t\ell,t'\ell'}(q_0)-\frac{1}{4\pi}\,\delta_{\ell\ell'}\delta_{tt'}
\mbox{Re}\left[M_\ell(q_0)\right]\, .
\en
The matrix that enters in the quantization condition, takes the form
\eq\label{eq:qc}
\mathscr{A}^\Gamma_{t\ell,t'\ell'}(q_0)
=\frac{1}{4\pi}\,\delta_{\ell\ell'}\delta_{tt'}K_\ell^M(q_0^2)-H^\Gamma_{t\ell,t'\ell'}(q_0)\, .
\en
Using Eq.~(\ref{eq:modified}) and the unitarity relation, which relates
the imaginary part of $M_\ell(q_0)$ with the Jost function $f_\ell(q_0)$, the matrix
$\mathscr{A}$ can be finally rewritten in the following form
\eq
\mathscr{A}^\Gamma_{t\ell,t'\ell'}(q_0)
=\frac{q_0^{2\ell+1}}{4\pi|f_\ell(q_0)|^2}\,\delta_{\ell\ell'}\delta_{tt'}
\cot(\delta_\ell(q_0)-\sigma_\ell(q_0))-\Delta H^\Gamma_{t\ell,t'\ell'}(q_0)\, .
\en
The advantage of this transformation consists in the fact that
the quantity $\Delta H$ does not contain ultraviolet divergences. Indeed, let us consider
the quantity $G_L(q_0)$, which enters the expression of the matrix $H$. The free Green
function in the finite-volume version of the
Lippmann-Schwinger equation~(\ref{eq:G_L}) can be
split into two parts, according to
\eq
\frac{1}{\bm{p}^2-q_0^2}=\underbrace{\left(\frac{1}{\bm{p}^2+\mu^2}
+\frac{\mu^2+q_0^2}{(\bm{p}^2+\mu^2)^2}+\cdots
+\frac{(\mu^2+q_0^2)^n}{(\bm{p}^2+\mu^2)^{n+1}}\right)}_{=G_1}
+\underbrace{\frac{(\mu^2+q_0^2)^{n+1}}{(\bm{p}^2+\mu^2)^{n+1}(\bm{p}^2-q_0^2)}}_{=G_2}\, .
\en
Here, $\mu$ is an arbitrary scale, and
the number of the subthreshold subtractions $n$ is taken large enough to ensure
the ultraviolet convergence of all sums after the angular-momentum truncation, when
$G_L$ is replaced by $G_2$. Furthermore, defining the scattering amplitude $T_1$
through the Lippmann-Schwinger equation $T_1=V_L+V_LG_1T_1$, we get
\eq
G_L=\underbrace{G_1+G_1T_1G_1}_{=G_\mu}
+\underbrace{(1+G_1T_1)\tilde G(T_1G_1+1)}_{=G_f}\, ,
\en
where
\eq
\tilde G=G_2+G_2T_1\tilde G\, .
\en
The infinite-volume limit in the above equations are easily performed by replacing
sums by integrals. Then, we have
\eq
\Delta H^\Gamma_{t\ell,t'\ell'}(q_0)&=&\frac{4\pi}{L^6}\,\sum_{\bm{p},\bm{q}}
\left(\mathscr{Y}^{\Gamma t}_\ell(\bm{p})\right)^*
\langle\bm{p}|G_\mu(q_0)+G_f(q_0)|\bm{q}\rangle
\mathscr{Y}^{\Gamma t'}_{\ell'}(\bm{q})
\nonumber\\[2mm]
&-&4\pi\int\frac{d^3\bm{p}}{(2\pi)^3}\,\frac{d^3\bm{q}}{(2\pi)^3}\,
\left(\mathscr{Y}^{\Gamma t}_\ell(\bm{p})\right)^*
\langle\bm{p}|G_\mu^\infty(q_0)+G_f^\infty(q_0)|\bm{q}\rangle
\mathscr{Y}^{\Gamma t'}_{\ell'}(\bm{q})\, .
\en
The key property of this expression is that the quantity $G_\mu$ does not have a singular
denominator in the physical region. Consequently, $G_\mu-G_\mu^\infty$ is suppressed
by factors of $\exp(-\mu L)$ and can be safely neglected.\footnote{A warning is in order:
  for certain values of the parameters, the quantities $G_\mu$ and $G_f$ can develop subthreshold poles that cancel in the sum. However, one can always adjust the free parameter $\mu$, so that these poles lie far outside the region of interest.} After dropping
$G_\mu-G_\mu^\infty$, the remainder contains the subtracted propagator $G_f$ only
and is ultraviolet-finite both in a finite and in the infinite volume separately. In this manner,
the quantity $\Delta H$ can be straightforwardly evaluated using numerical methods.
Ultraviolet divergences are absent. No partial-wave expansion is involved for the
long-range potential, and the Lippmann-Schwinger equation in a finite volume is solved
in a plane-wave basis.

\begin{figure}[t]
  \begin{center}
\includegraphics*[width=0.35\paperwidth]{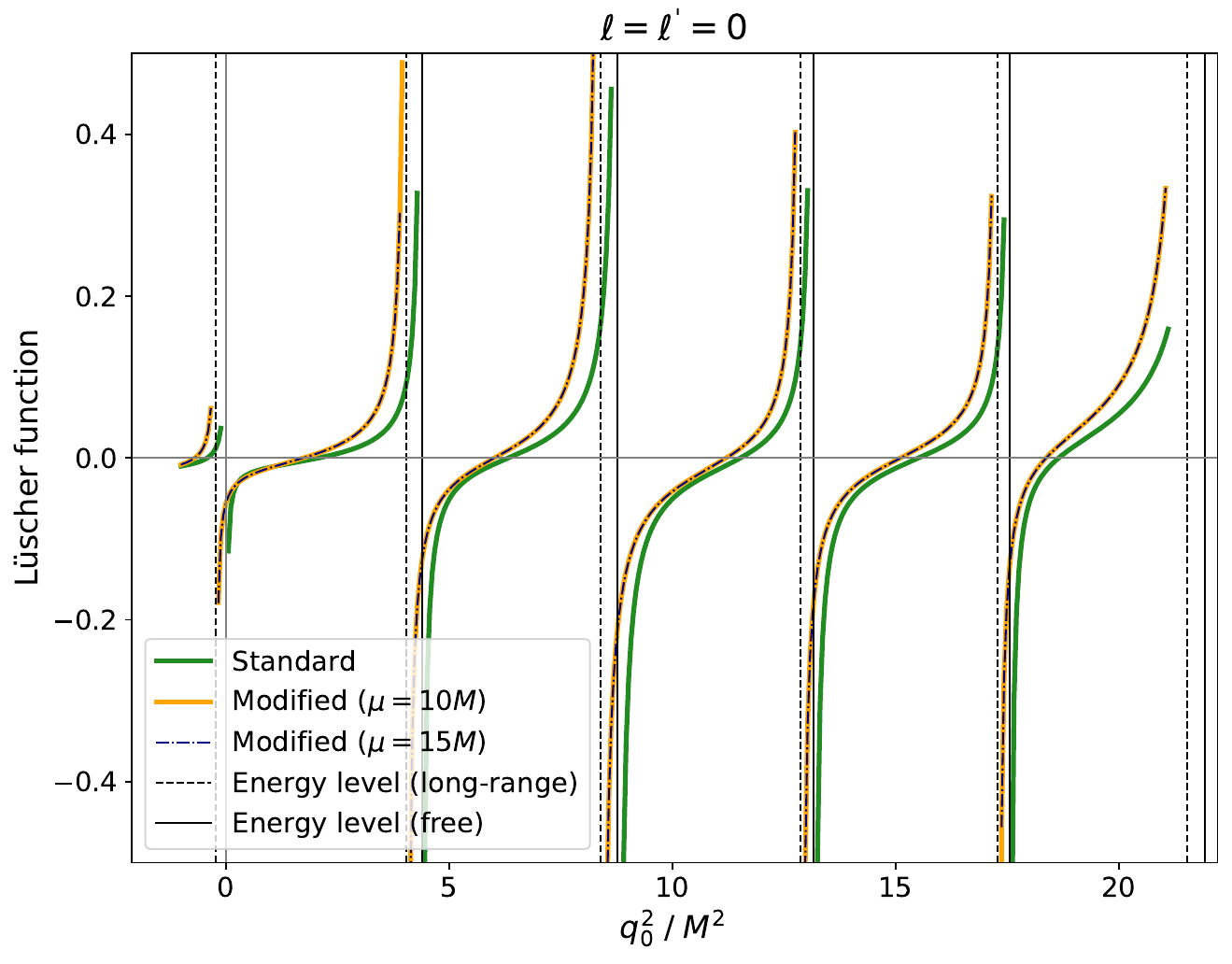}
\includegraphics*[width=0.35\paperwidth]{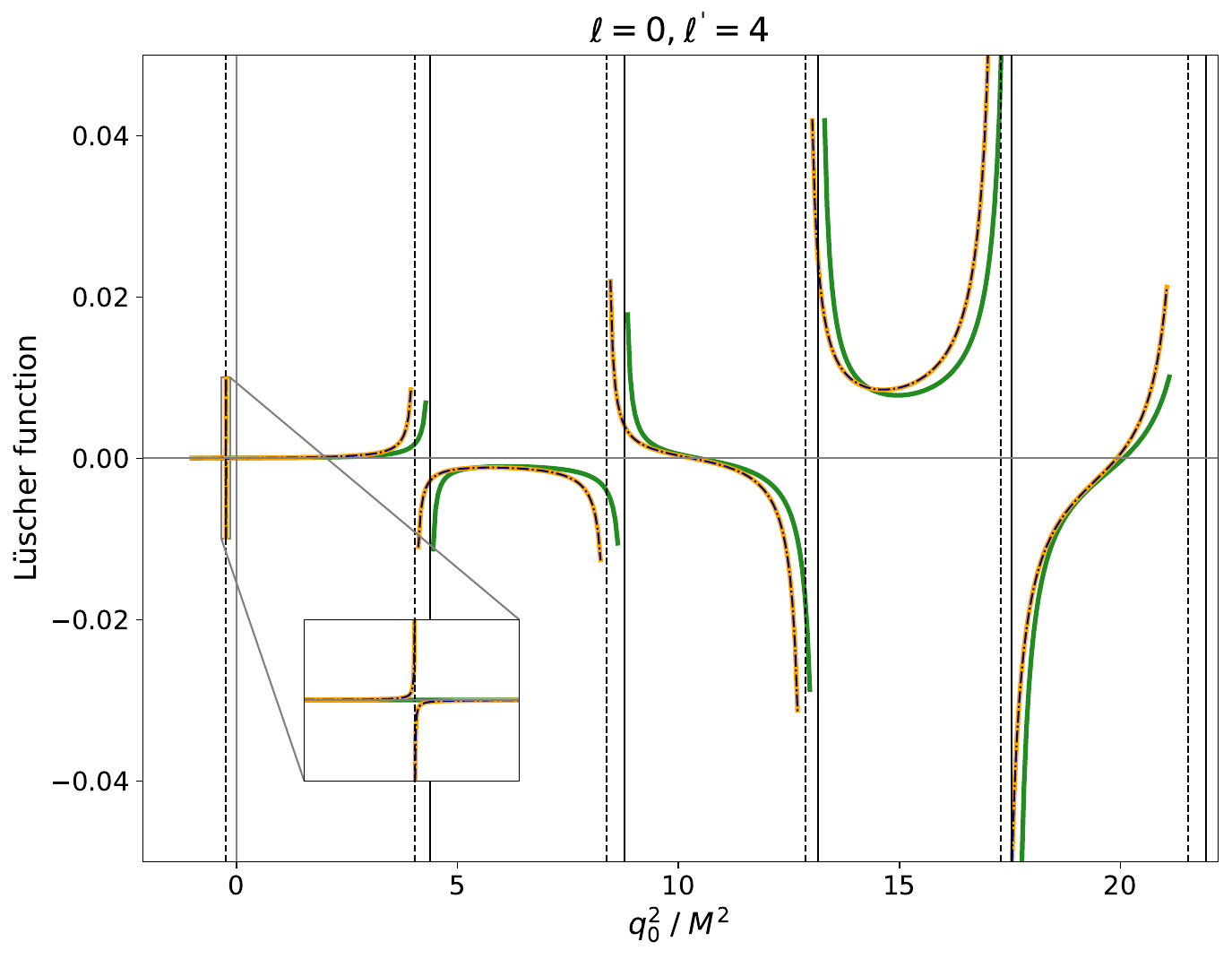}
\\
\includegraphics*[width=0.35\paperwidth]{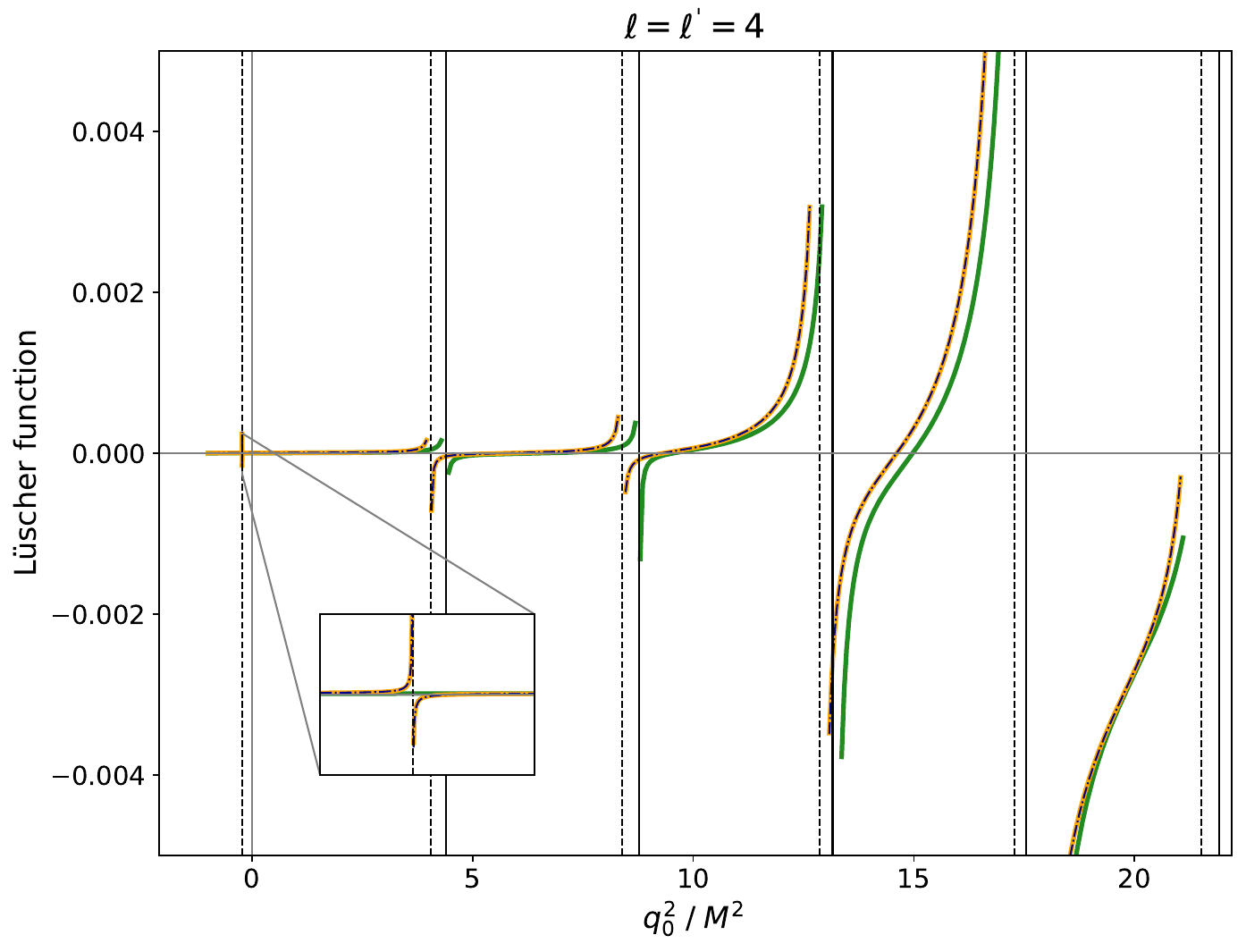}
\caption{Modified L\"uscher zeta-function vs. standard L\"uscher zeta-function for two
  different values of parameter $\mu$: $\mu=10M$ and $\mu=15M$. The size of the
  box is fixed by $ML=3$.
As seen, the difference between these two values of $\mu$ is not seen by a bare eye.
  Furthermore, the vicinity of the lowest energy level is zoomed in the last two plots, where the modified function, unlike the standard one, has a pole.}
\label{fig:mu_indept}
\end {center}
\end{figure}

In Fig.~\ref{fig:mu_indept} we plot the modified L\"uscher zeta-function\footnote{Strictly speaking,  $H^\Gamma_{t\ell,t'\ell'}(q_0)$ in the free case is given by a linear combination of the zeta-functions with known group-theoretical coefficients. For brevity, we shall refer to  $H^\Gamma_{t\ell,t'\ell'}(q_0)$ as to the zeta-function as well.} $H^\Gamma_{t\ell,t'\ell'}(q_0)$ (see Eq.~(\ref{eq:modified_qc})) vs. the standard L\"uscher zeta-function obtained
from $H^\Gamma_{t\ell,t'\ell'}(q_0)$ at $g=0$. The angular momenta take the values
$\ell,\ell'=0,4$, and $q_0^2=E/M$. The labels $t,t'$ are neglected, because multiple representations are not present. The calculations of the modified L\"uscher zeta-function are done
for two different values of the parameter $\mu$. As seen from Fig.~\ref{fig:mu_indept}, the results are practically independent of $\mu$, which serves as a very good check on the calculations.

The calculated modified and standard functions are very similar to each other,
except for a small horizontal shift. A difference, however, emerges in the vicinity of the lowest
energy level corresponding to the relative momentum $\bm{p}=(0,0,0)$:
except in the case $\ell=\ell'=0$, the standard L\"uscher zeta-function does not have a pole
here. However, the modified function is a solution of the Lippmann-Schwinger type
equation, where all momenta are present in the intermediate state, and therefore it
develops a pole at the energy corresponding to the lowest level. For a better visibility,
the singular behavior of the modified L\"uscher zeta-function in this energy region is zoomed
in for the cases $\ell=0,\ell'=4$ and $\ell=\ell'=4$.

\subsection{Analysis of the synthetic data}

The potential of the toy model, which are used to produce the synthetic lattice data, is
given by a sum of the long- and short-range parts:
\eq\label{eq:potential_toy}
V(\bm{p},\bm{q})=\frac{4\pi g}{M^2+(\bm{p}-\bm{q})^2}
+\frac{4\pi g_S}{M_S^2+(\bm{p}-\bm{q})^2}\, ,
\en
with $g_S/g=-2.74$. In addition, we set the ratio $m/M=6.7$. Moreover, we treat
$M_S/M$ as a free parameter, in order to observe a smooth transition from the
short-range to the long-range interaction.

First of all, note that one can easily solve the finite-volume quantization condition with the
above potential, finding the eigenvalues of the Hamiltonian in the plane-wave basis.
These solutions will be used to test the validity of our method, since both approaches coincide up to the exponentially small corrections. Note also that, in the Hamiltonian approach, one can mimic the truncation of higher partial waves  in the L\"uscher equation. To this end, one has merely to project the input potential onto the partial waves and drop higher partial waves in this projection.

As we know, one of the fundamental problems in using L\"uscher equation in the presence
of a long-range force consists in a slow convergence of the partial-wave expansion.
In order to compare the convergence speed in case of the standard/modified L\"uscher
approach, we retain only S- and G-waves in the quantization condition. The equation
(\ref{eq:qc}) then yields 
\eq\label{eq:DE}
D(E)=\left(\frac{1}{4\pi}\,K_0^M(q_0)-H_{00}(q_0)\right)-\frac{(H_{04}(q_0))^2}{\left(\dfrac{1}{4\pi}\,K_4^M(q_0)-H_{44}(q_0)\right)}=0\, .
\en
Here the labels $t,t'$ are omitted for brevity. In case of the standard L\"uscher equation, $K_\ell^M(q_0)$ is replaced
by $K_\ell(q_0)$, and $H_{\ell\ell'}(q_0)$ by the expression evaluated at $g=0$. If only the
S-wave is taken into account, the second term in the expression (\ref{eq:DE})
is dropped.
Calculations are done for two values $M_S/M=10$ and $M_S/M=2$, corresponding
to the two extreme cases. 

\begin{figure}[t]
  \begin{center}
 \includegraphics*[width=0.35\paperwidth]{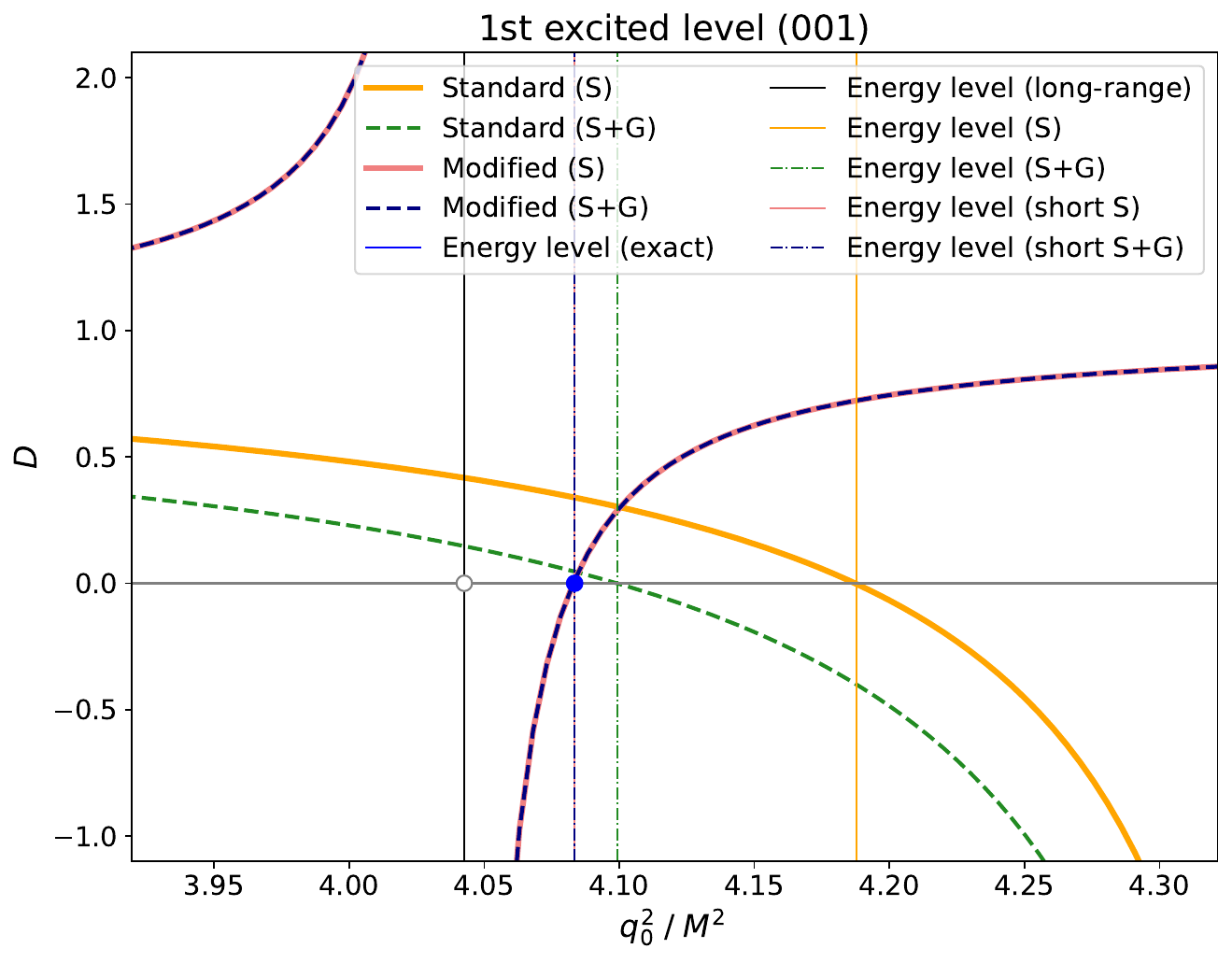}
 \includegraphics*[width=0.35\paperwidth]{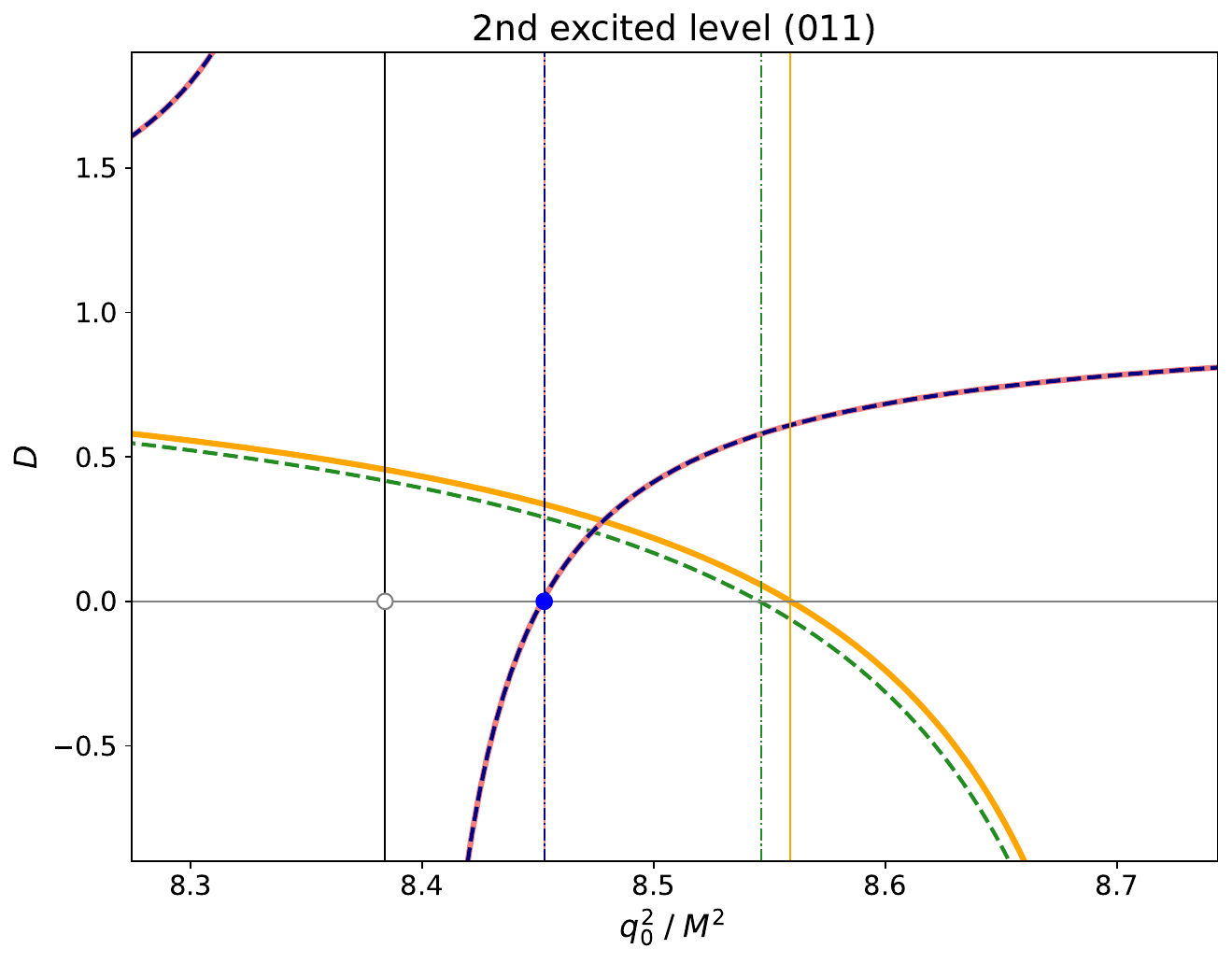}
\\
\includegraphics*[width=0.35\paperwidth]{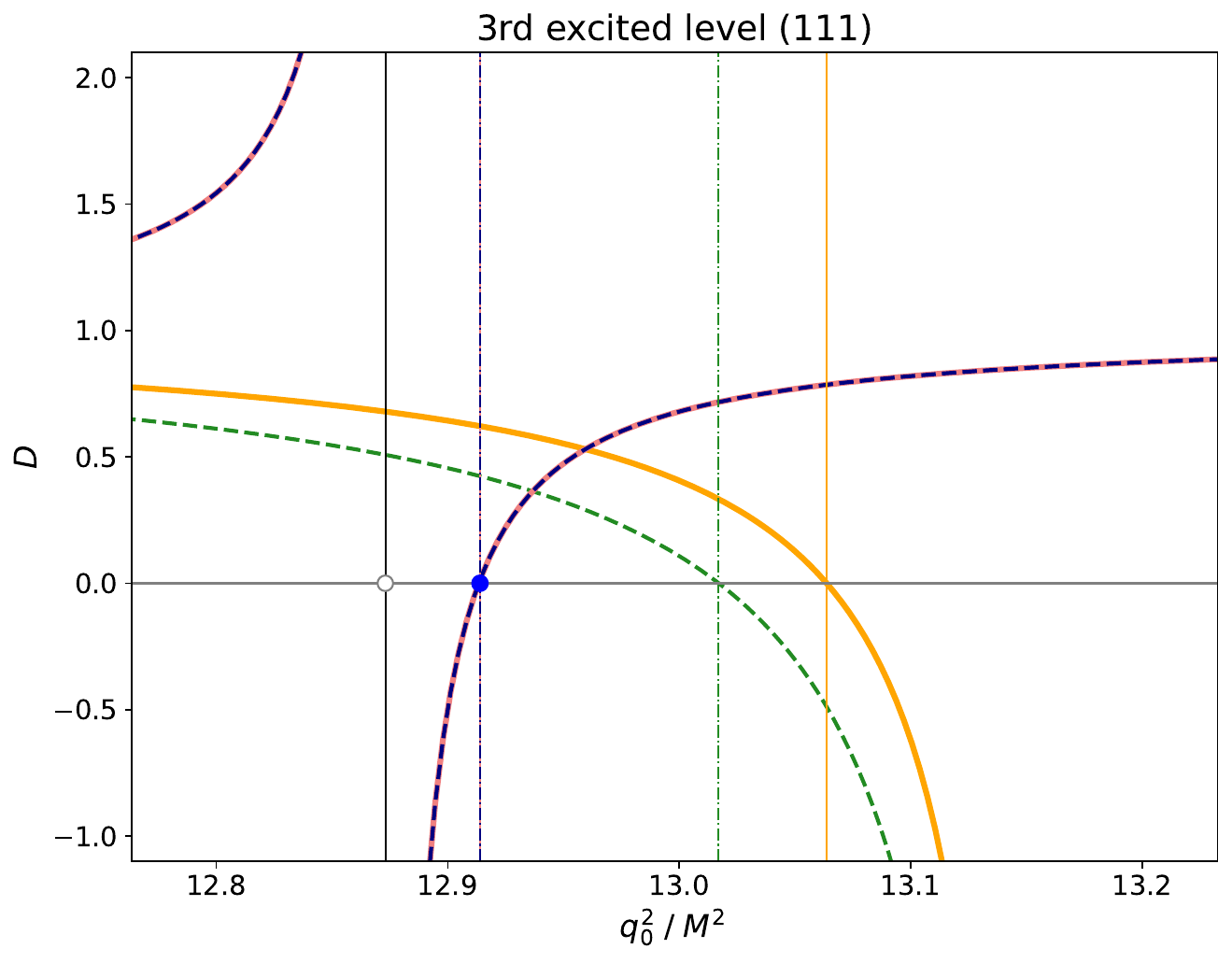}
\includegraphics*[width=0.35\paperwidth]{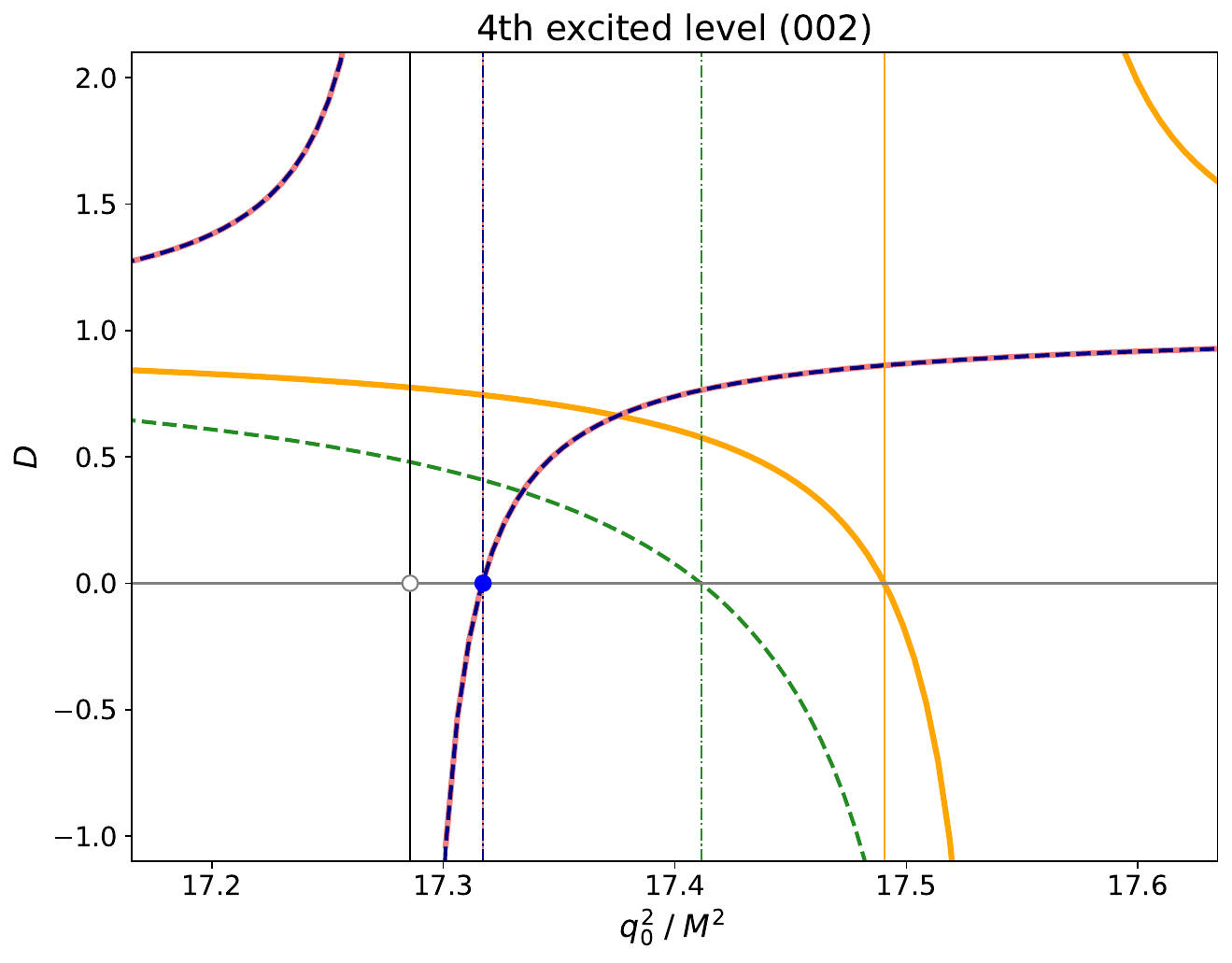}
\caption{The solution of the quantization condition for the first few excited levels
  (both the standard and modified L\"uscher equation). The partial-wave expansion is truncated, retaining only S-wave, or S- and G-waves. For comparison, the eigenvalues of the
  Hamiltonian are shown by vertical lines: exact, projected on S, S+G waves, only the short-range part projected on S, S+G waves. In addition, we show a solution with the long-range potential only, i.e., $V_S=0$. The values of the parameters used are 
  $M_S=10M$, $ML=3$. In the title of each plot, we indicate an unperturbed plane-wave state momentum, to which this state reduces in the absence of the interaction.}
\label{fig:detA_10}
\end {center}
\end{figure}

Let us start with the excited states. The solutions of the quantization condition for the
first few excited levels and for $M_S/M=10$ are seen in Fig.~\ref{fig:detA_10}.
This plot beautifully demonstrates the advantage of the modified quantization condition
with respect to the standard one. Namely, while taking into account the partial waves
with $\ell=0,4,\ldots$, the solution of the standard L\"uscher equation slowly converges
to the exact solution. In contrast, the solution of the modified quantization condition with the
S-wave only already reproduces the exact energy level, and adding the G-wave does not
change anything -- the two curves cannot be distinguished from each other.

\begin{figure}[t]
  \begin{center}
 \includegraphics*[width=0.35\paperwidth]{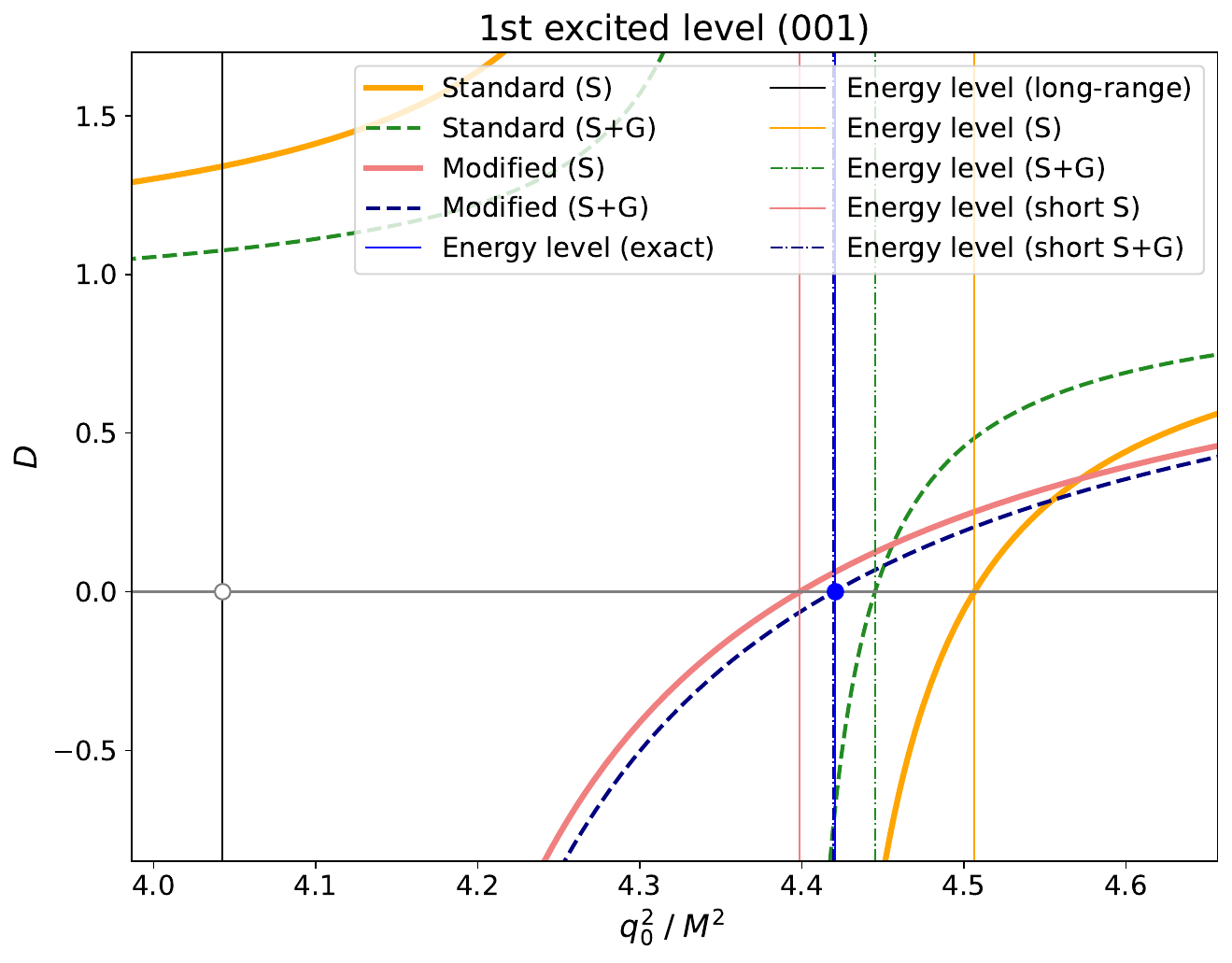}
 \includegraphics*[width=0.35\paperwidth]{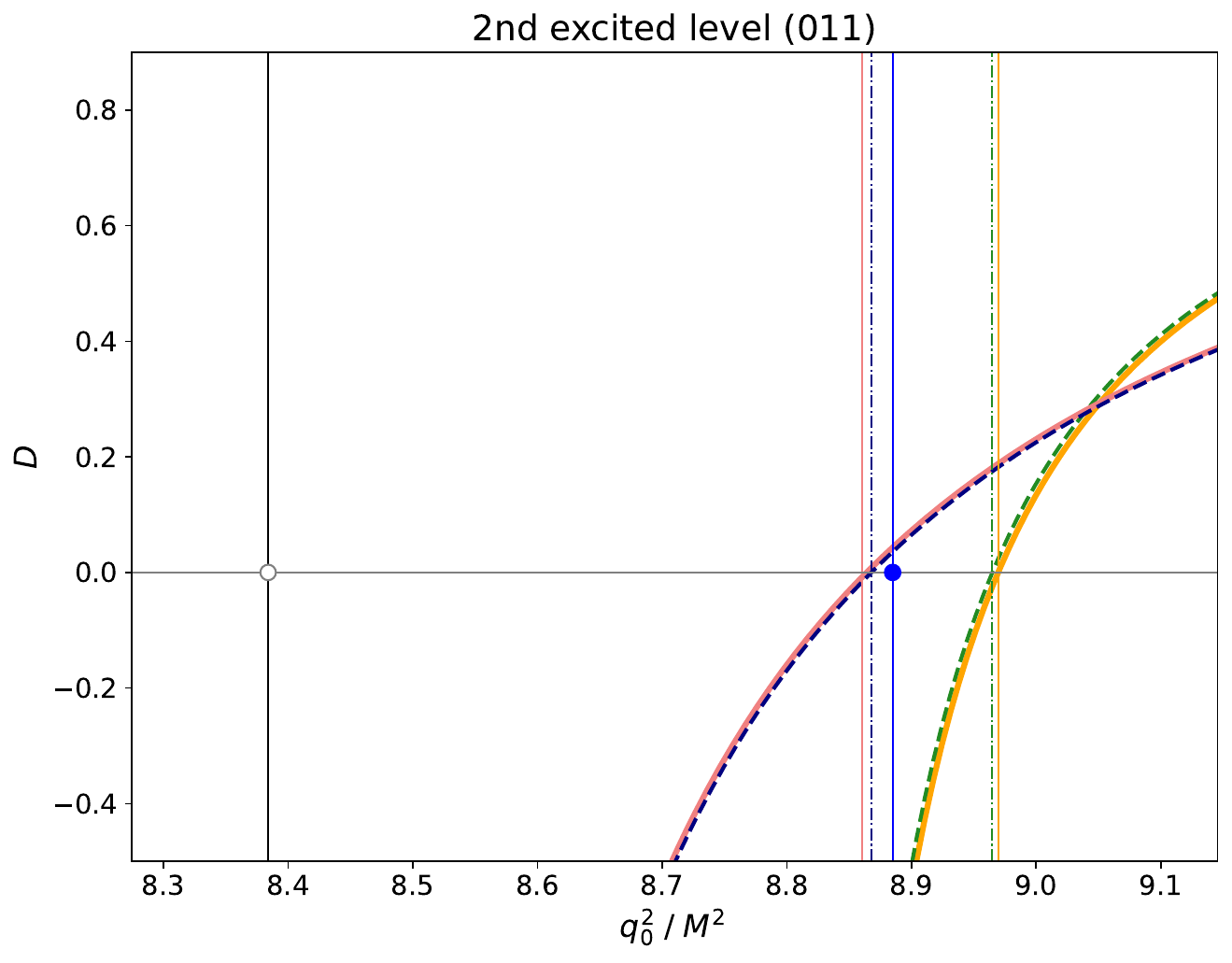}
\\
\includegraphics*[width=0.35\paperwidth]{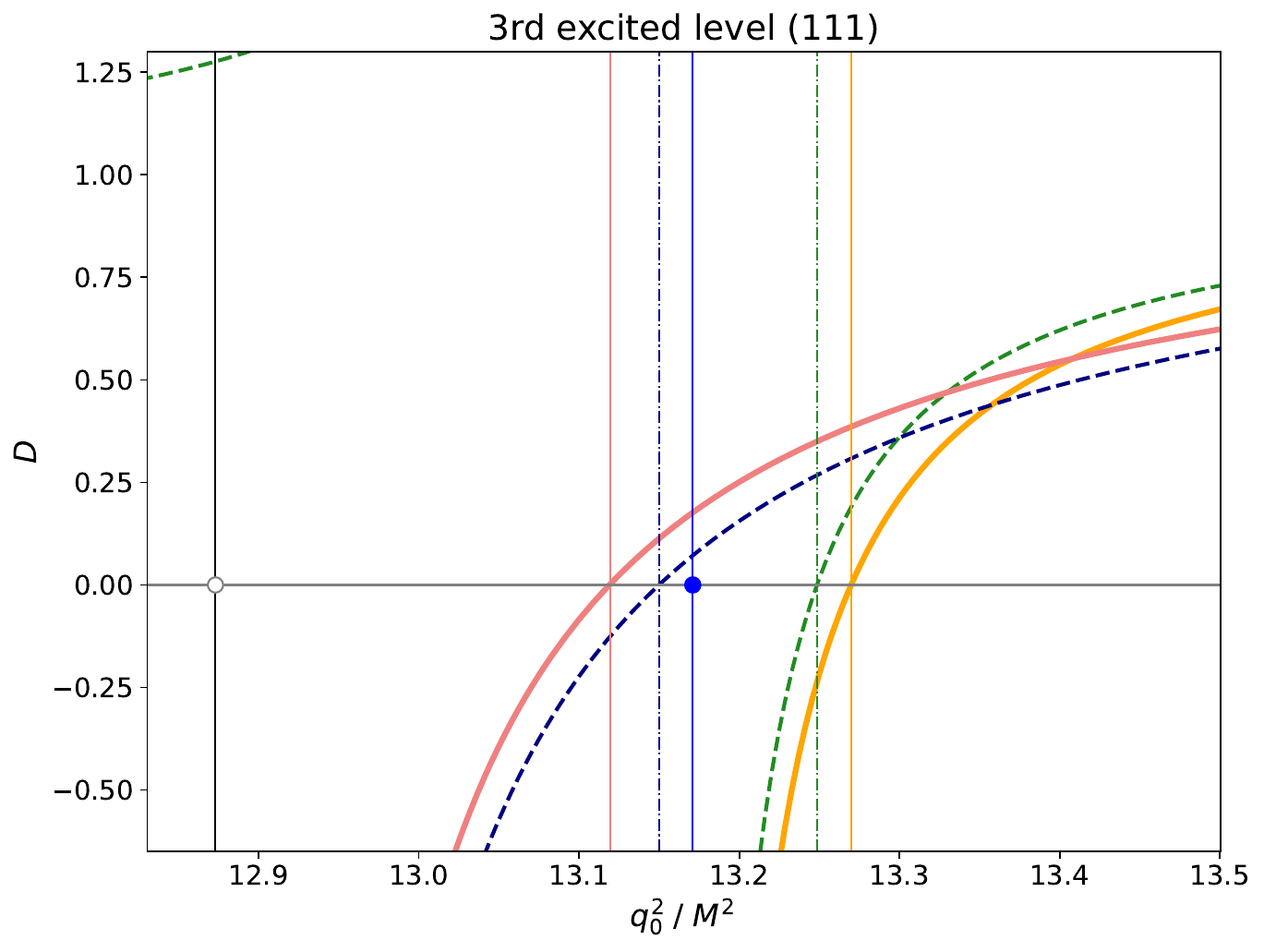}
\includegraphics*[width=0.35\paperwidth]{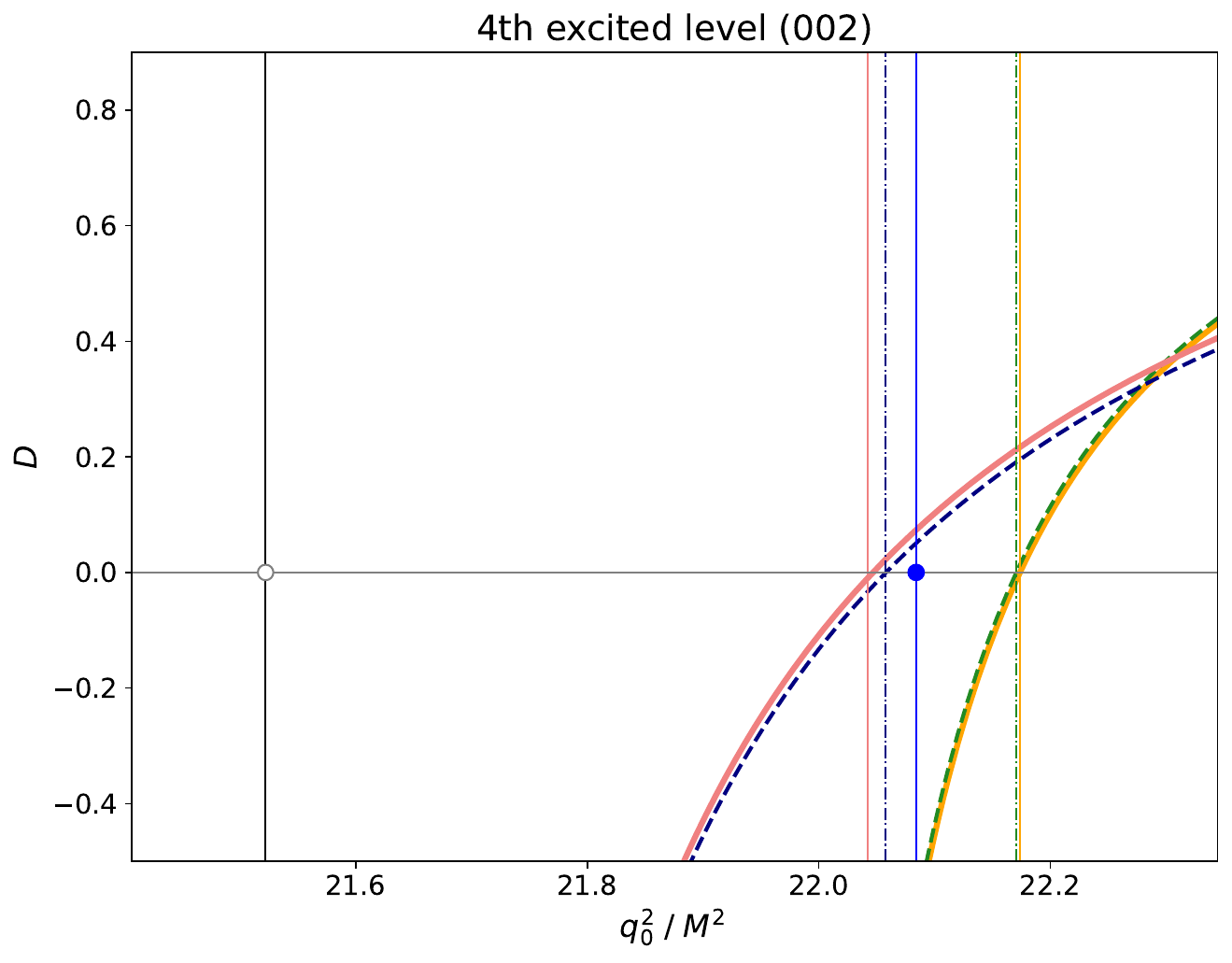}
\caption{The same as in Fig.~\ref{fig:detA_10}, but for $M_S=2M$.}
\label{fig:detA_2}
\end {center}
\end{figure}

The pattern, described above, remains practically unchanged in case of $M_S/M=2$, shown in Fig.~\ref{fig:detA_2},
which describes a borderline situation with $V_S$ almost long-range. Now, adding the
G-wave leads to a visible effect, but still the convergence of the modified quantization condition is much better as compared to the standard one.

\begin{figure}[t]
  \begin{center}
 \includegraphics*[width=0.15\paperwidth]{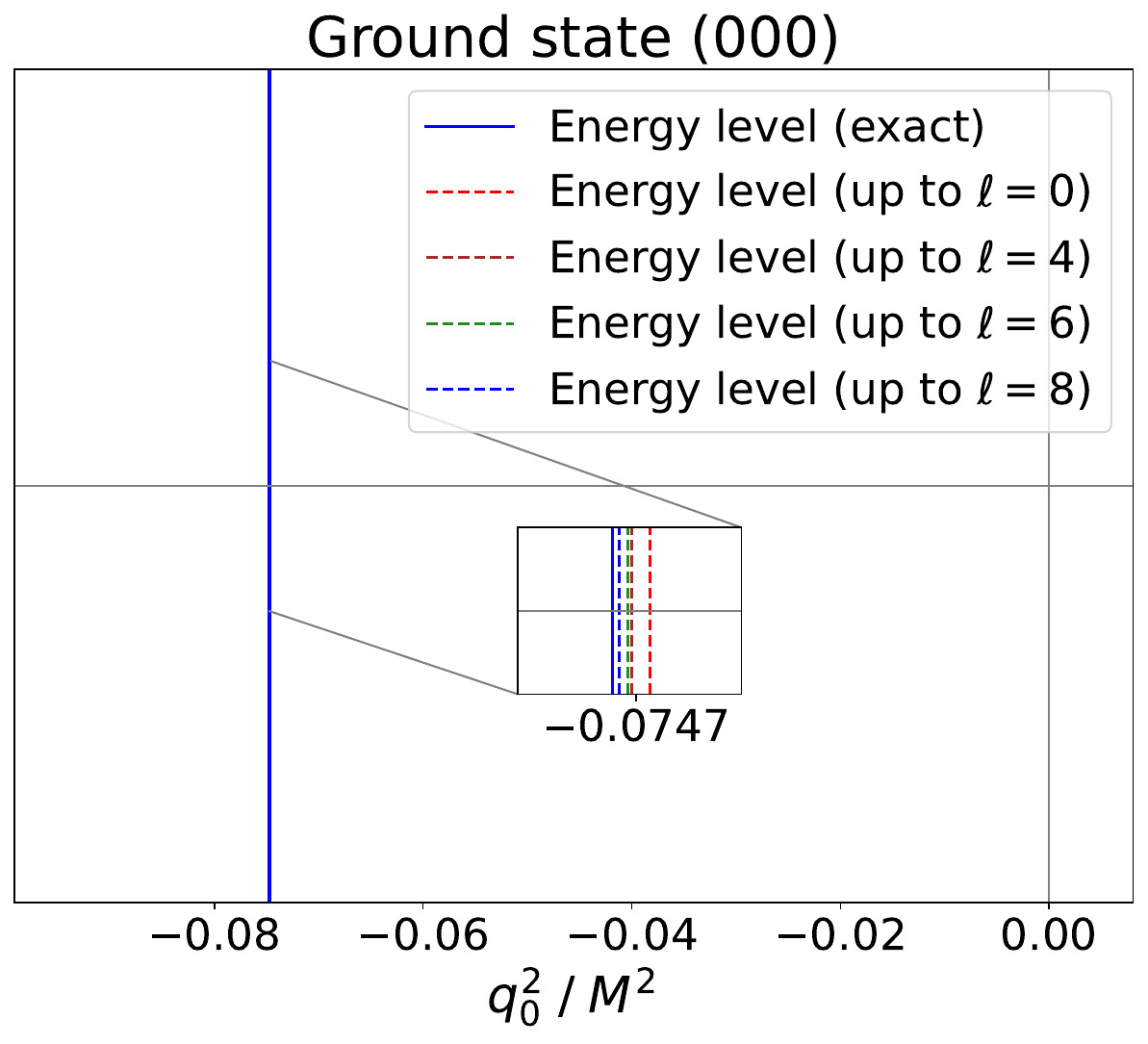}
 \includegraphics*[width=0.15\paperwidth]{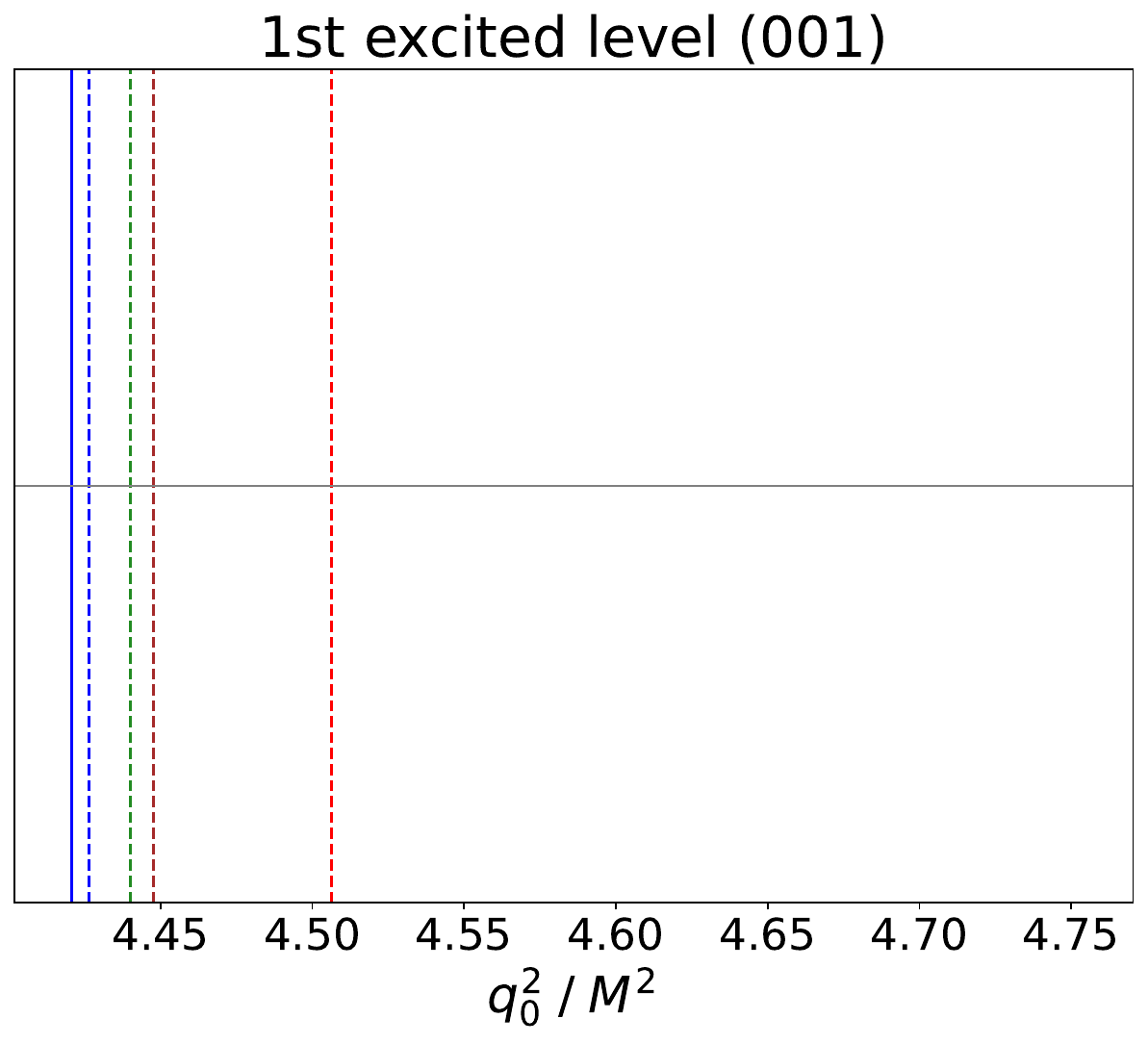}
 \includegraphics*[width=0.15\paperwidth]{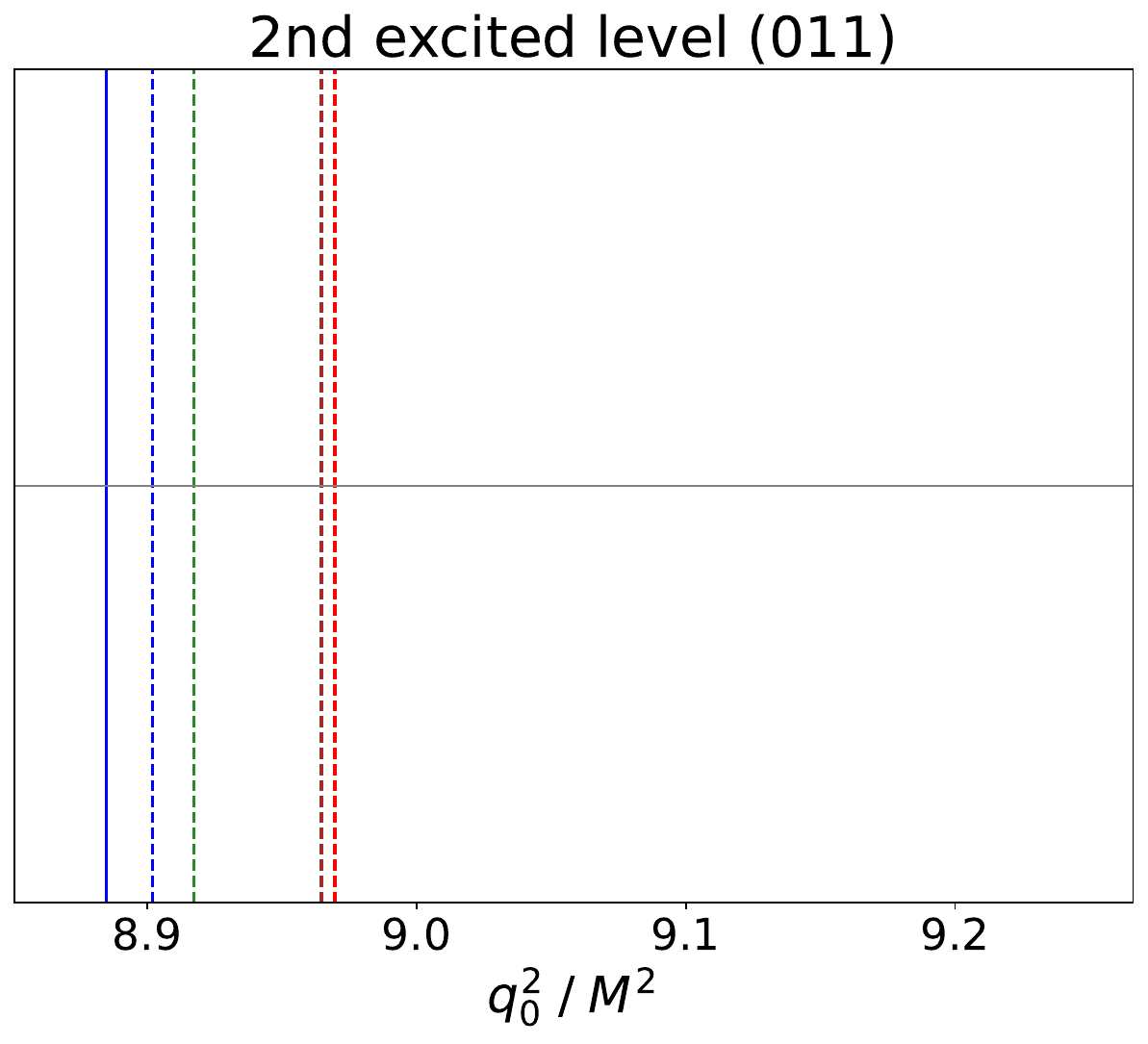}
 \includegraphics*[width=0.15\paperwidth]{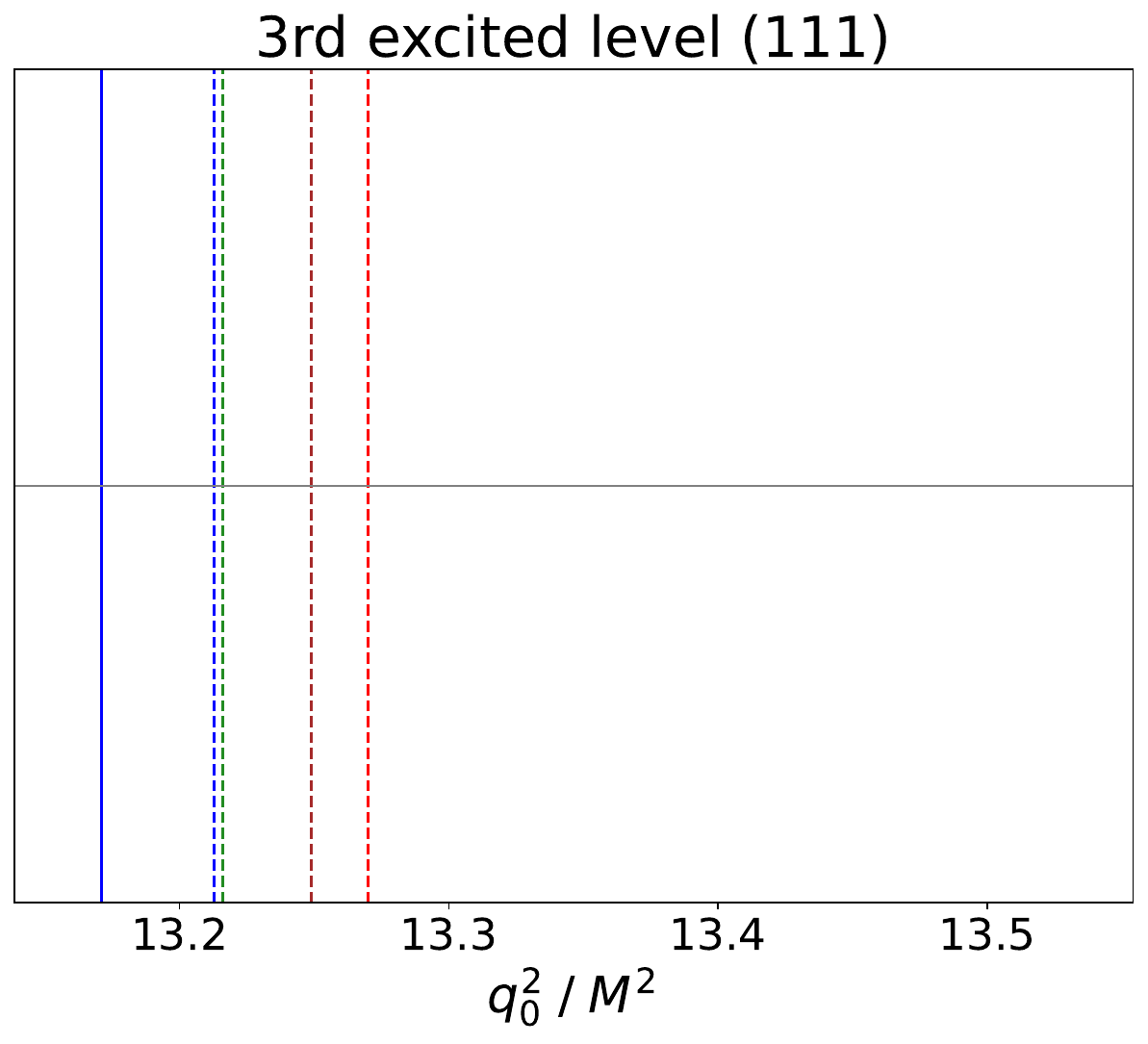}
\\
 \includegraphics*[width=0.15\paperwidth]{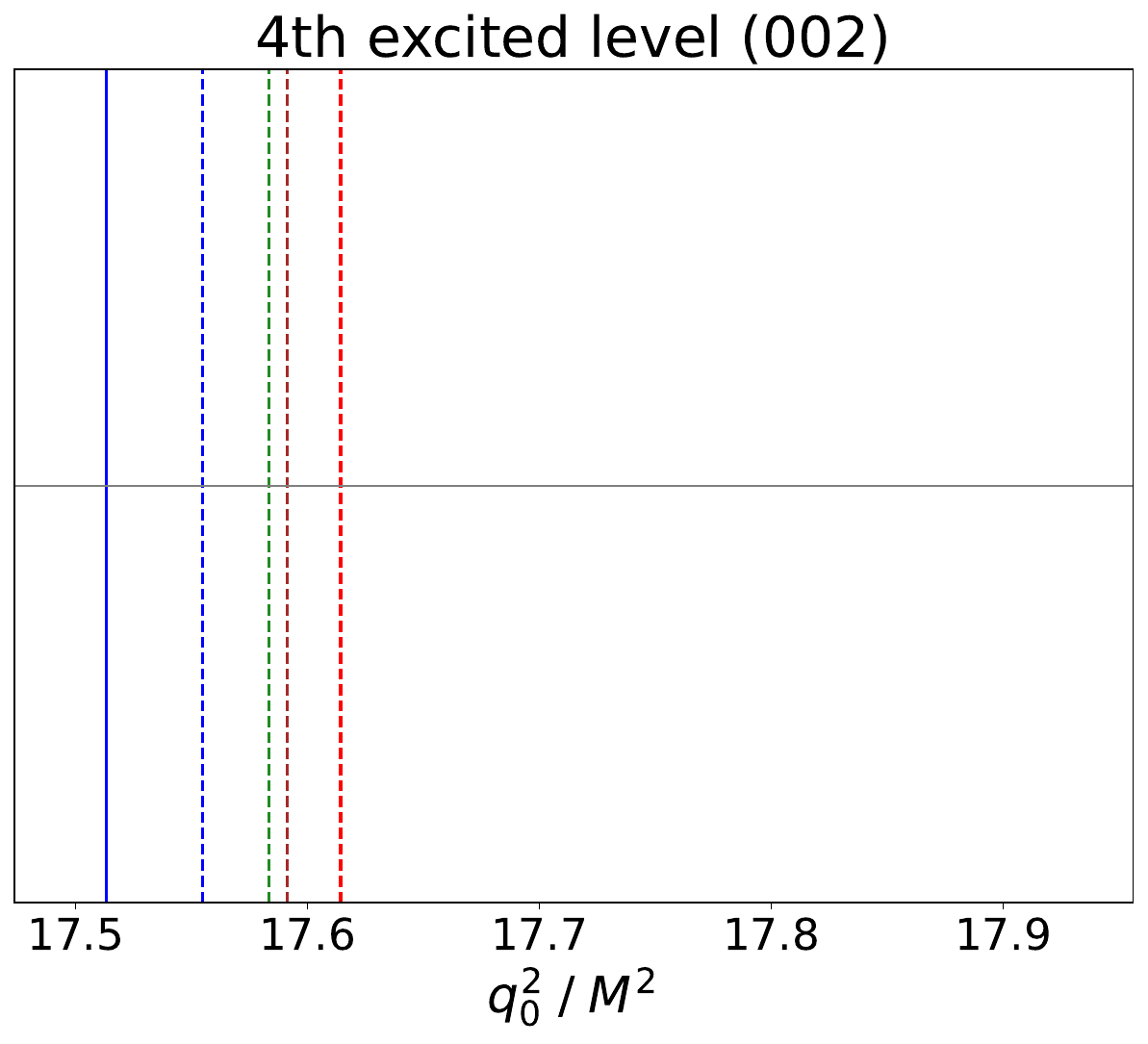}
 \includegraphics*[width=0.15\paperwidth]{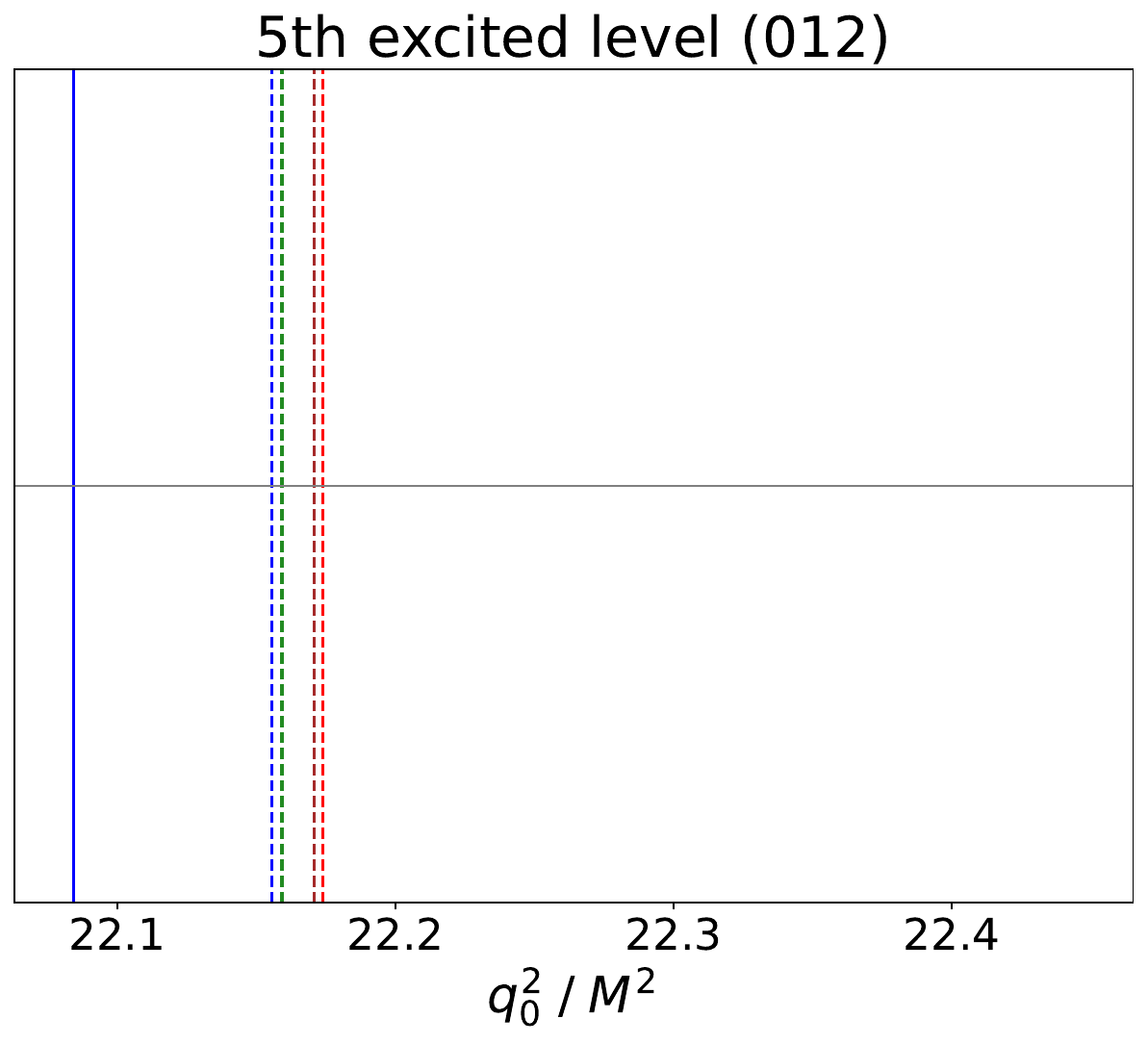}
 \includegraphics*[width=0.15\paperwidth]{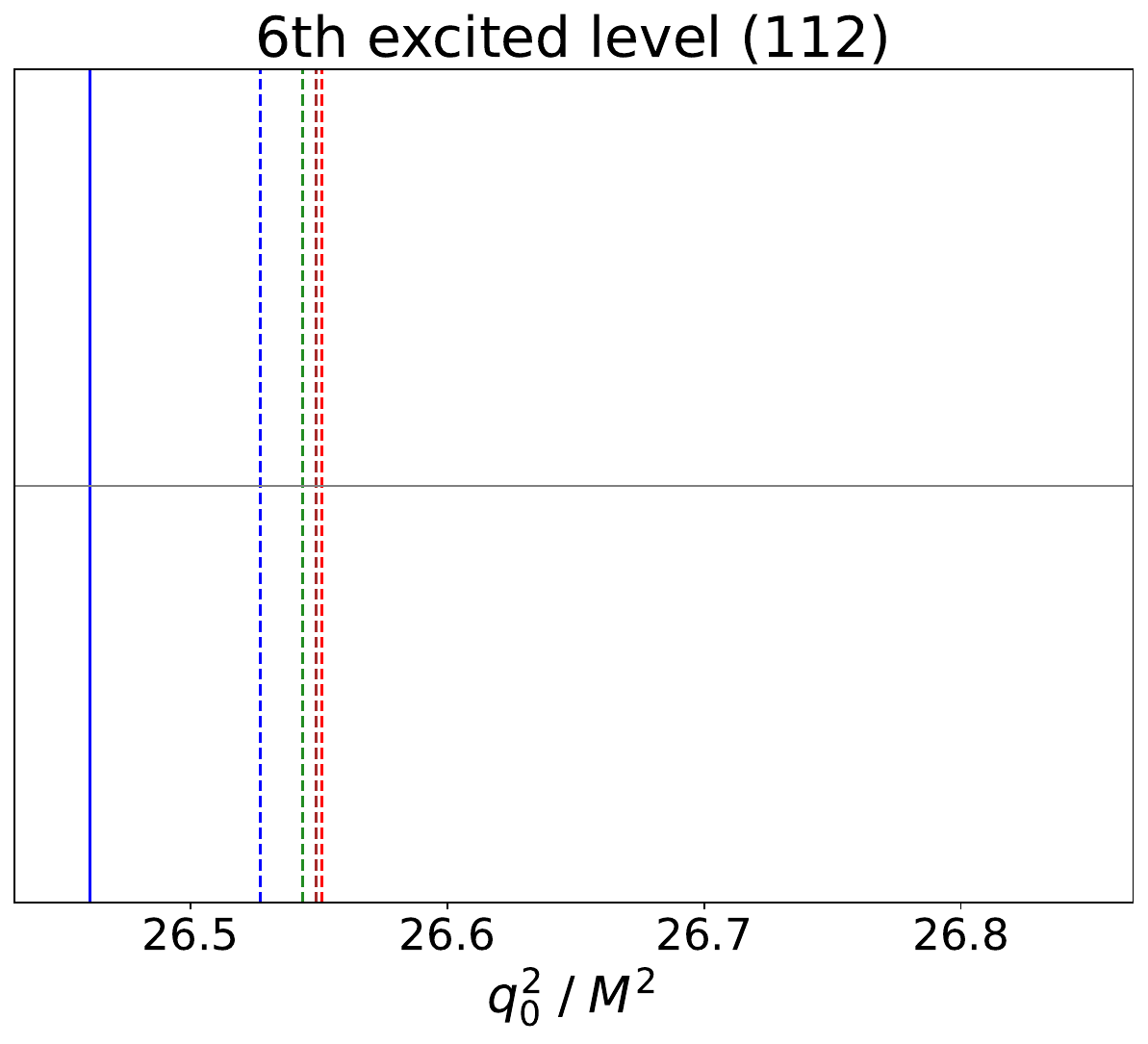}
 \includegraphics*[width=0.15\paperwidth]{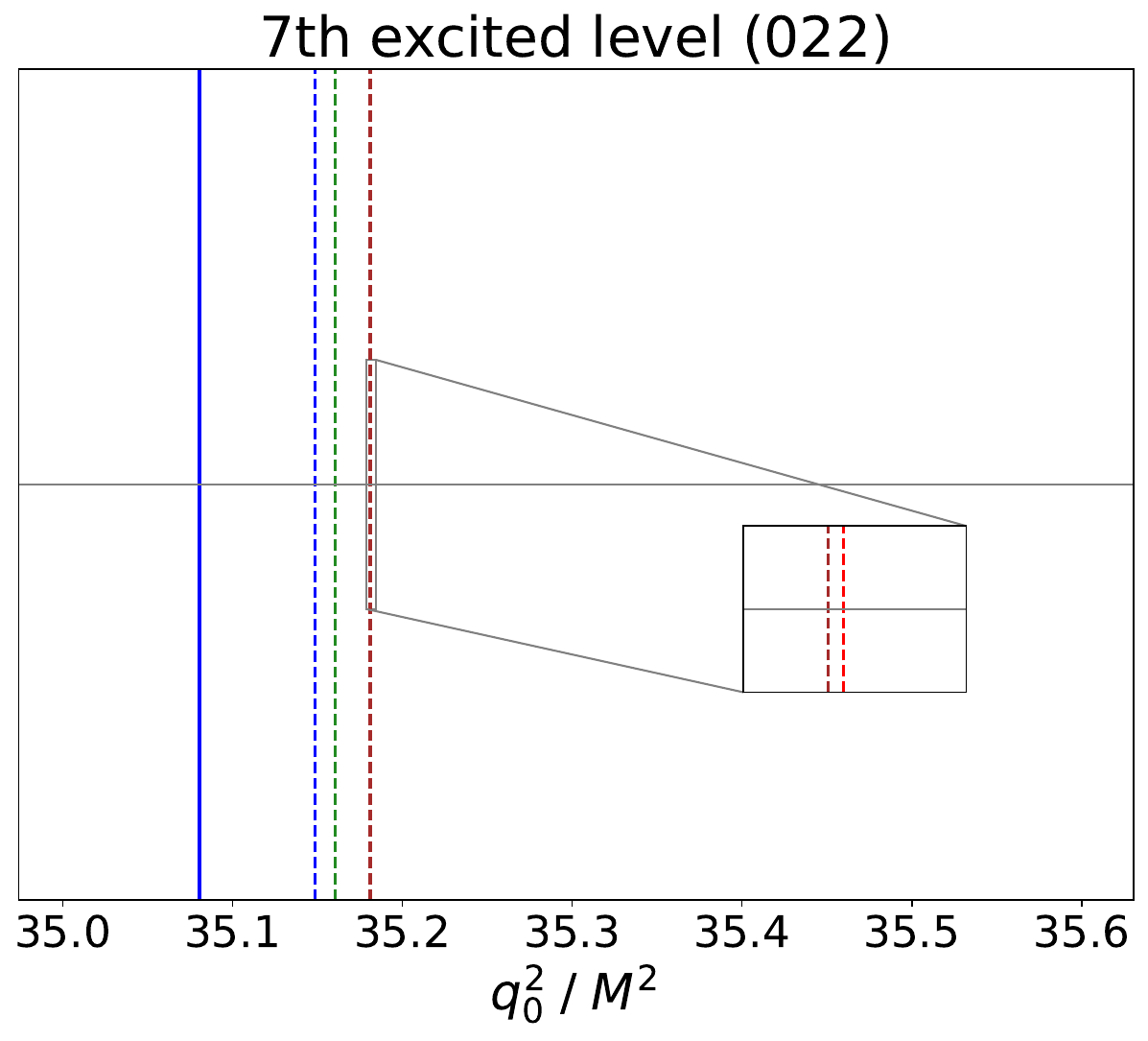}
 \caption{Checking the convergence of the partial-wave expansion in the energy spectrum
   of the Hamiltonian. The values of the parameters used are $M_S=10M$ and $ML=3$.}
\label{fig:HEFT}
\end {center}
\end{figure}

Few comments are in order. First, as seen, the exponentially suppressed effects are
very small, because the solutions of the quantization condition reproduce very well the
finite-volume spectrum of the Hamiltonian in the plane-wave basis. Second, the
convergence of the partial-wave expansion is rather uneven. This can be seen better
from Fig.~\ref{fig:HEFT}, which shows the finite-volume spectrum of the Hamiltonian with
the full potential projected onto different partial waves. As seen, for example, the contribution of the G-wave to the 2nd excited state is rather small, and the higher partial waves contribute significantly. A qualitative discussion of this phenomenon is given in appendix~\ref{app:PW}. In general, as expected, the convergence becomes slower for higher partial waves, with relative momentum increased.

Last but not least, we would like to note that in the situation when the short-range is
completely absent (i.e., when $V_S=0$), the energy levels are given by the poles of
the modified L\"uscher zeta-function $H_{\ell\ell'}(q_0)$. In Figs.~\ref{fig:detA_10} and \ref{fig:detA_2}, these poles are indicated by empty circles. 
It is seen that, even for $M_S=10M$, the effect of setting $V_S=0$ is sizable. However,the S-wave contribution alone suffices to reproduce the exact answer.
This means that the partial-wave expansion in the modified L\"uscher equation converges
very fast indeed, and the truncation to the S-wave yields a very good approximation,
in difference to the standard approach.

\begin{figure}[t]
  \begin{center}
 \includegraphics*[width=0.35\paperwidth]{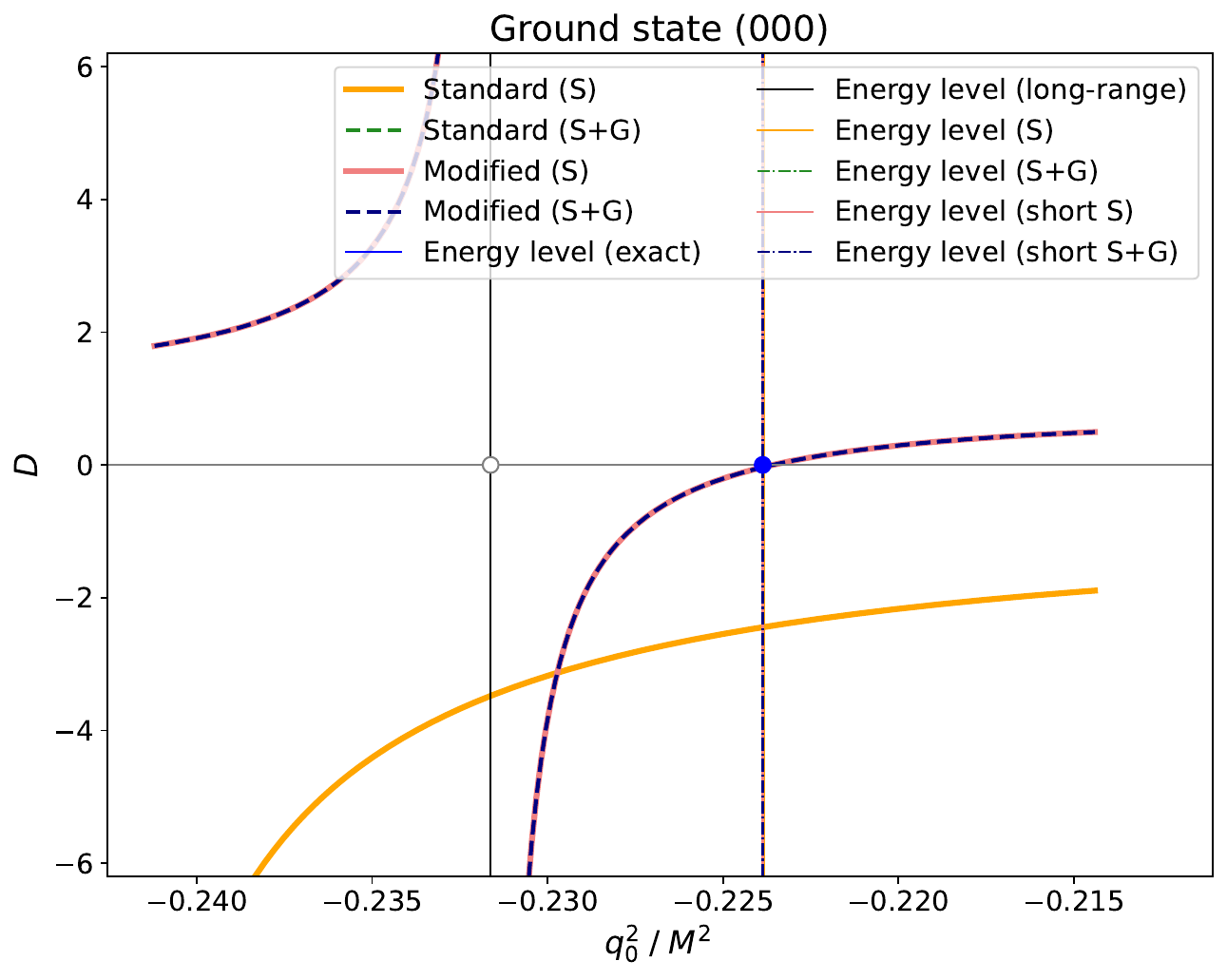}
 \includegraphics*[width=0.35\paperwidth]{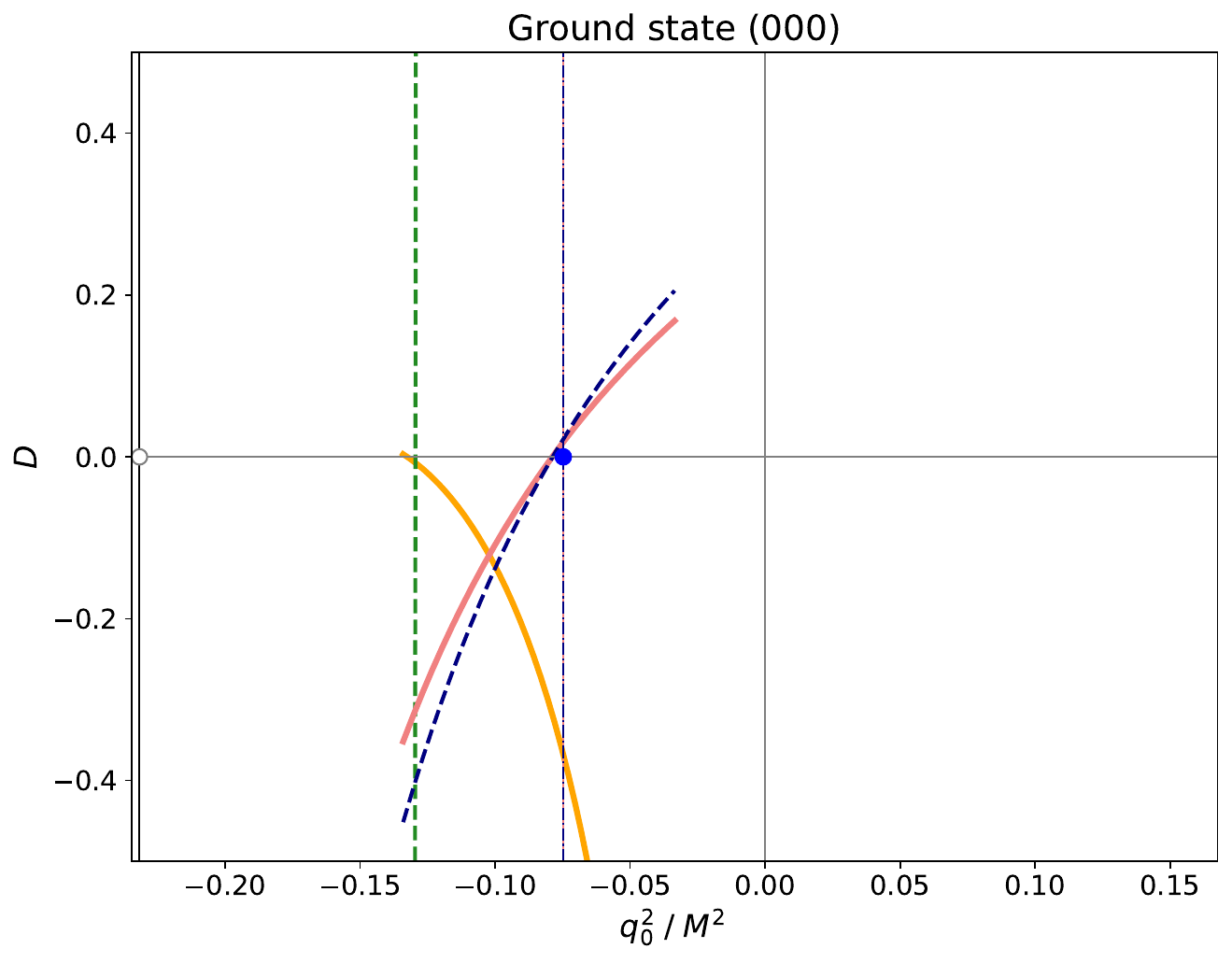}
 \caption{The solution of the quantization condition (both standard and modified)
   for the ground state. The values of the parameters are $ML=3$ and
   $M_S=10M$ (left panel), $M_S=2M$ (right panel). There is no solution of the standard quantization condition in case $M_S=10M$.}
\label{fig:detA_ground}
\end {center}
\end{figure}

The ground state is considered separately. First, note that the convergence
of the partial-wave expansion in the Hamiltonian approach is very good\footnote{ It is worth to mention that for some particular excited state, the $G$-wave contribution is obvious suppressed, for example, the 2nd excited level which is around the momentum of $(0,1,1)$. A detailed discussion is presented in App.~\ref{app:PW}.}, see Fig.~\ref{fig:HEFT}, even for the standard case. The solution of the quantization condition, however, shown in Fig.~\ref{fig:detA_ground} presents different picture.
The modified quantization condition still provides a very accurate solution at $M_S/M=10$
and a quite accurate solution at $M_S/M=2$. The standard L\"uscher equation, however, is completely off. The reason for this is the proximity of the left-hand cut which leads to large exponential corrections. As expected, these corrections are strongly suppressed
in the modified quantization condition as compared to the standard one.

In order to understand this result qualitatively, let us introduce a characteristic parameter that determines, how far the left-hand cut is located from a given energy level. In the case of standard and modified L\"uscher equations, the left hand starts at $q_0^2/M^2=-1/4$ and $q_0^2/M^2=-M_S^2/4M^2$, respectively. If one divides this quantity by the value of $q_0^2/M^2$ corresponding to the ground state (taken from the figure) and extracts the square root, one ends up with a ratio of momentum scales. In the case of the standard L\"uscher equation,
this ratio is equal to $1.1$ and $1.9$ for $M_S=10M$ and $M_S=2M$, respectively. In the modified case, this ratio becomes $11$ and
$3.8$, respectively. This, in our opinion, explains why the exponential corrections in the modified case are much smaller than in the standard setting. It would be interesting to see, which value a similar parameter takes in the analysis of data carried out in Ref.~\cite{Collins:2024sfi}, since this might provide a rough estimate of the exponential corrections which are neglected in this analysis.

\section{Modified effective range expansion in the left-hand cut region}
\label{sec:dimensional}

\subsection{Calculation of the loop function $M_\ell(q_0)$}

The function $M_\ell(q_0)$ is given by Eq.~(\ref{eq:M_ell}). As already mentioned, this
quantity is ultraviolet-divergent. One way to regularize this divergence is to render the
long-range potential superregular, e.g., by using the Pauli-Villars regularization or a
sharp cutoff. The price to pay for using this straightforward procedure is that the
coefficients of the modified effective range expansion if the function $K_\ell^M(q_0^2)$
are not of natural size anymore. Indeed, let $\Lambda$ denote the (large) cutoff mass
used in the regularization. For a given $\ell$, the loop diverges as $\Lambda^{2\ell+1}$.
Since in the modified effective range expansion this loop enters in a combination
$K_\ell^M(q_0^2)-M_\ell(q_0)$, it is clear that this polynomial with large coefficients has
to be absorbed into $K_\ell^M(q_0^2)$. At the practical level, the cutoff regularization
becomes numerically inconvenient for even not so large values of $\ell$.\footnote{It
  should be mentioned that the problem of the ultraviolet divergences in the modified effective range expansion has been observed in the literature before,
  see, e.g., Ref.~\cite{Minossi}.} For this reason, we propose to use dimensional regularization
instead, where the power divergences are absent. However, in this case, we encounter
another problem. Namely, the quantity $G_L(q_0)$, which enters the expression for
$M_\ell(q_0)$, contains an infinite sum of one-pion exchange ladder diagrams
(see Fig.~\ref{fig:loops}) and is
determined through the solution of an integral equation. It is not obvious how one can
use dimensional regularization to tame divergences in the integral equation.

The key observation that allows one to solve the problem is that each consequent
term in the Born series of $G_L(q_0)$ has a lower index of divergence than the previous
one. Therefore, only a finite number of terms in the expansion diverge. One can use
this property and split $G_L(q_0)$ into two parts: $G_L(q_0)=G_L^{\sf div}(q_0)
+G_0(q_0)T_L^{\sf fin}G_0(q_0)$, and, correspondingly, $M_\ell(q_0)=M_\ell^{\sf div}(q_0)+
M_\ell^{\sf fin}(q_0)$,
where
\eq\label{eq:div-fin}
G_L^{\sf div}(q_0)&=&\underbrace{\left(G_0(q_0)+G_0(q_0)V_LG_0(q_0)+\cdots\right)}_{2\ell+2~\mbox{terms}}\, ,
\nonumber\\[2mm]
T_L^{\sf fin}&=&\underbrace{\left(V_L+V_LG_0(q_0)V_L+\cdots\right)}_{2\ell+2~\mbox{terms}}
+V_LG_0(q_0)T_L^{\sf fin}\, .
\en
The integral equation that determines $T_L^{\sf fin}$ is of a Lippmann-Schwinger type and
can be numerically solved by using standard techniques. The loop integral, containing
$T_L^{\sf fin}$, is free of ultraviolet divergences. Hence all calculations can be carried out
in $d=3$ dimensions.

The remainder is given by the following expression:
\eq
M_\ell^{\sf div}(q_0)=4\pi\int\frac{d^3\bm{p}}{(2\pi)^3}\,\frac{d^3\bm{q}}{(2\pi)^3}\,
(pq)^\ell P_\ell(\hat p\cdot\hat q)\langle\bm{p}|G_L^{\sf div}(q_0)|\bm{q}\rangle\, .
\en
In order to carry out dimensional regularization in the above expression, one should
first define Legendre polynomials $P_\ell(x)$ in $d$ dimensions. This definition is not
unique. However, using different definitions reduces to adding a finite polynomial in
$q_0^2$ to $M_\ell^{\sf div}(q_0)$ and hence can be accounted for by adjusting the
renormalization prescription. Our definition is based on the use of the $d$-dimensional
harmonic polynomials $\mathscr{Y}^{(\ell)}_{i_1\cdots i_\ell}(\bm{p})
=y_{i_1\cdots i_\ell,j_1\cdots j_\ell}p_{j_1}\cdots p_{j_\ell}$, which have the following
properties:
\begin{itemize}
\item
  $\mathscr{Y}^{(\ell)}_{i_1\cdots i_\ell}(\bm{p})$ is a polynomial in the components of the vector $\bm{p}$;
\item
  $\mathscr{Y}^{(\ell)}_{i_1\cdots i_\ell}(\bm{p})$ is symmetric and traceless in any pair of indices.
\end{itemize}
The examples of harmonic polynomials for $\ell=0,1,2,3,4$ are given below:
\eq
\mathscr{Y}^{(0)}(\bm{p})&=&1\, ,
\nonumber\\[2mm]
\mathscr{Y}^{(1)}_i(\bm{p})&=&p_i\, ,
\nonumber\\[2mm]
\mathscr{Y}^{(2)}_{ij}(\bm{p})&=&p_ip_j-\frac{1}{d}\,\bm{p}^2\, ,
\nonumber\\[2mm]
\mathscr{Y}^{(3)}_{ijk}(\bm{p})&=&p_ip_jp_k-\frac{1}{d+2}\,\bm{p}^2(\delta_{ij}p_k
+\delta_{ik}p_j+\delta_{jk}p_i)\, ,
\nonumber\\[2mm]
\mathscr{Y}^{(4)}_{ijkl}(\bm{p})&=&p_ip_jp_kp_l-\frac{1}{d+4}\,\bm{p}^2
(\delta_{ij}p_kp_l+\delta_{ik}p_jp_l+\delta_{il}p_jp_k+\delta_{jk}p_ip_l+\delta_{jl}p_ip_k
+\delta_{kl}p_ip_j)
\nonumber\\[2mm]
&&\quad\quad\quad~\,+\,\frac{1}{(d+2)(d+4)}\, \bm{p}^4(\delta_{ij}\delta_{kl}+\delta_{ik}\delta_{jl}+\delta_{il}\delta_{jk})\, ,
\en
and so on. The coefficients $y_{i_1\cdots i_\ell,j_1\cdots j_\ell}$ can be easily read off from
these expressions.
Furthermore,
\eq
(pq)^\ell P_\ell(\hat p\cdot\hat q)=c_d^{(\ell)}\sum_{i_i,\cdots,i_\ell}\mathscr{Y}^{(\ell)}_{i_1\cdots i_\ell}(\bm{p})\mathscr{Y}^{(\ell)}_{i_1\cdots i_\ell}(\bm{q})\, .
\en
This gives
\eq
P_0(\hat p\cdot\hat q)&=&1\, ,
\nonumber\\[2mm]
(pq)P_1(\hat p\cdot\hat q)&=&(\bm{p}\bm{q})\, ,
\nonumber\\[2mm]
(pq)^2P_2(\hat p\cdot\hat q)&=&\frac{d}{d-1}\,\left((\bm{p}\bm{q})^2-\frac{1}{d}\,\bm{p}^2\bm{q}^2\right)\, ,
\nonumber\\[2mm]
(pq)^3P_3(\hat p\cdot\hat q)&=&\frac{d+2}{d-1}\,\left((\bm{p}\bm{q})^3
  -\frac{3}{d+2}\,(\bm{p}\bm{q})\bm{p}^2\bm{q}^2\right)\, ,
\nonumber\\[2mm]
(pq)^4P_4(\hat p\cdot\hat q)&=&\frac{(d+2)(d+4)}{d^2-1}\,
\left((\bm{p}\bm{q})^4-\frac{6}{d+4}\,(\bm{p}\bm{q})^2\bm{p}^2\bm{q}^2
+\frac{3}{(d+2)(d+4)}\,\bm{p}^4\bm{q}^4\right)\, ,\quad\quad
\en
and so on. The normalization constant $c_d^{(\ell)}$ is chosen so that
$P_\ell(1)=1$ for all $d$.

Since our calculations are restricted to the $A_1^+$ irrep of the octahedral group
for demonstration purposes, it suffices to consider the angular momenta $\ell=0,4$.

\subsubsection{The case $\ell=0$}

In this case, only the two-loop diagram $M_{0,2}^{\sf div}(q_0)$
(the second diagram in Fig.~\ref{fig:loops}) should be considered. 
The one-loop diagram with no pion exchange can be easily evaluated in dimensional
regularization  and the rest is ultraviolet-finite. The two-loop diagram in dimensional
regularization is given by
\eq
M_{0,2}^{\sf div}(q_0)&=&4\pi\int \frac{d^d\bm{p}}{(2\pi)^d}\,
\frac{d^d\bm{q}}{(2\pi)^d}\,\frac{1}{\bm{p}^2-q_0^2}\,\frac{4\pi g}{M^2+(\bm{p}-\bm{q})^2}\,\frac{1}{\bm{q}^2-q_0^2}
\nonumber\\[2mm]
&=&\int_0^1\mathscr{D}_{xy}\int \frac{d^d\bm{p}}{(2\pi)^d}\,\frac{d^d\bm{q}}{(2\pi)^d}\,\frac{\Gamma(3)\,(4\pi)^2g}{\left(x_1(\bm{p}^2-q_0^2)+y(M^2+(\bm{p}-\bm{q})^2)
  +x_2(\bm{q}^2-q_0^2)\right)^3}\, ,\quad\quad
\en
where
\eq
\mathscr{D}_{xy}=dx_1dx_2dy\,\delta(1-x_1-x_2-y)\, .
\en
One can rewrite the above expression in a compact form, introducing a $2$-dimensional
column $\bm{Q}_\alpha,\,\alpha=1,2$, where $\bm{Q}_1=\bm{p}$ and
$\bm{Q}_2=\bm{q}$. Then,
\eq\displaystyle
M_{0,2}^{\sf div}(q_0)=\int_0^1\mathscr{D}_{xy}\int \frac{d^{2d}\bm{Q}}{(2\pi)^{2d}}\,\dfrac{\Gamma(3)\,(4\pi)^2g}{\left(\sum\limits_{\alpha,\beta}
    \bm{Q}_\alpha C_{\alpha\beta}\bm{Q}_\beta
    +\Delta\right)^3}\, ,
\en
where
\eq
C=\begin{pmatrix}
  x_1+y & -y \cr
  -y & y+x_2
\end{pmatrix}\, ,\quad\quad
\Delta=yM^2-(x_1+x_2)q_0^2\, .
\en
One can now perform orthogonal transformations in order to diagonalize the matrix $C$.
The result looks as
\eq\displaystyle
M_{0,2}^{\sf div}(q_0)=\int_0^1\mathscr{D}_{xy}\int \frac{d^{2d}\bm{Q}}{(2\pi)^{2d}}\,\dfrac{\Gamma(3)\,(4\pi)^2g}{\left(\sum\limits_\alpha
    \bm{Q}_\alpha \lambda_\alpha\bm{Q}_\alpha
    +\Delta\right)^3}\, ,
\en
where $\lambda_\alpha$ denote the eigenvalues of the matrix $C$. Rescaling further the momenta $\bm{Q}_\alpha\to\lambda_\alpha^{-1/2}\bm{Q}_\alpha$, we get:
\eq\label{eq:ell=0}\displaystyle
M_{0,2}^{\sf div}(q_0)&=&\int_0^1\mathscr{D}_{xy}\int \frac{d^{2d}\bm{Q}}{(2\pi)^{2d}}\,\dfrac{\Gamma(3)\,(4\pi)^2g\det(C)^{-d/2}}{\left(\sum\limits_\alpha
    \bm{Q}_\alpha \bm{Q}_\alpha
    +\Delta\right)^3}
\nonumber\\[2mm]
&=&\frac{\Gamma(3-d)g}{(4\pi)^{d-2}}\,\int_0^1\mathscr{D}_{xy}
\det(C)^{-d/2}\Delta^{d-3}\, .
\en
We further introduce the new variables $x_1=\rho\tau$ and $x_2=\rho(1-\tau)$,
with $0\leq\rho,\tau\leq 1$. Defining the dimensionless quantity $z=q_0^2/M^2$
and carrying out the integration in $y$, we get
\eq\displaystyle
M_{0,2}^{\sf div}(q_0)&=&
\frac{\Gamma(3-d)g(M^2)^{d-3}}{(4\pi)^{d-2}}\,\int_0^1
\rho d\rho d\tau((1-\rho)-\rho z)^{d-3}(\rho(1-\rho)+\rho^2\tau(1-\tau))^{-d/2}
\nonumber\\[2mm]
&=&\frac{\Gamma(3-d)g(M^2)^{d-3}}{(4\pi)^{d-2}}\,\int_0^\infty ds\int_0^1 d\tau
(s-z)^{d-3}(s+\tau(1-\tau))^{-d/2}
\nonumber\\[2mm]
&=&-\frac{g}{2(d-3)}+\mbox{UV finite},
\en
where a new integration variable $s=(1-\rho)/\rho$ was introduced.

As anticipated, the two-loop diagram is ultraviolet divergent. One could use, for example,
MS or $\overline{\mbox{MS}}$ renormalization scheme, using $M$
(as the only available mass) as a renormalization scale. The advantage of such a procedure
consists in the fact that no large polynomials should be absorbed in $K_\ell^M(q_0^2)$
anymore. It should be pointed, however, that using this scheme becomes more
complicated for higher values of $\ell$, even if the above statement still applies. Namely,
there are more divergent diagrams for higher $\ell$, and the factors $(d-3)^{-1}$ may
arise either from the prefactor containing the $\Gamma$-function, or the integral over the
variable $s$, and the separation of the all ultraviolet-divergent terms becomes involved.
We find that a simple solution to the problem is to impose a different renormalization
prescription, namely, to subtract first few terms in the Taylor expansion at threshold $z=0$. In the case we are considering, a single subtraction suffices, since the divergent term
is a constant:
\eq
\hat M_{0,2}^{\sf div}(q_0)=M_{0,2}^{\sf div}(q_0)-M_{0,2}^{\sf div}(0)\, .
\en
For higher $\ell$, more subtractions are necessary.

\subsubsection{The case $\ell=4$}

In the section above we have considered the case $\ell=0$ in great detail, including
the discussion of the ultraviolet regularization and the choice of the renormalization prescription.
Most of the techniques, used here, can be applied for higher values of $\ell$ without change.
In this case, however, there are more divergent diagrams.
The $N$-loop diagram can be written as follows: 
\eq
M_{4,N}^{\sf div}(q_0)&=&4\pi\int \frac{d^d\bm{Q}_1}{(2\pi)^d}\cdots
\frac{d^d\bm{Q_N}}{(2\pi)^d}\,\frac{1}{\bm{Q}_1^2-q_0^2}\,\frac{4\pi g}{M^2+(\bm{Q}_1-\bm{Q}_2)^2}\cdots\frac{(Q_1Q_N)^4P_4(\hat Q_1\cdot \hat Q_N)}{\bm{Q}_N^2-q_0^2}
\nonumber\\[2mm]
&=&\int_0^1\mathscr{D}_{xy}\int \frac{d^d\bm{Q}_1}{(2\pi)^d}\cdots\frac{d^d\bm{Q}_N}{(2\pi)^d}\,
\frac{h_N(Q_1Q_N)^4P_4(\hat Q_1\cdot \hat Q_N)}
{D^{2N-1}}\, ,\quad\quad\,\,
\en
where $h_N=\Gamma(2N-1)\,(4\pi)^Ng^{N-1}$ and
\eq
D=x_1(\bm{Q}_1^2-q_0^2)
+y_1(M^2+(\bm{Q}_1-\bm{Q}_2)^2)+\cdots
+x_N(\bm{Q}_N^2-q_0^2)\, .
\en
Here, the Feynman parameters $x_i,\,i=1,\ldots,N$ and $y_i,i=1,\ldots,N-1$ are associated
with the Green function $(\bm{Q}_i^2-q_0^2)^{-1}$ and the pion propagator
$(M^2+(\bm{Q}_i-\bm{Q}_{i+1})^2)^{-1}$, respectively, and
$\mathscr{D}_{xy}=dx_1\cdots dx_Ndy_1\cdots dy_{N-1}\delta(1-x_1-\cdots-x_N-y_1-\cdots-y_{N-1})$. The tridiagonal matrix $C$ takes the following form
\eq
C=\begin{pmatrix}
  x_1+y_1 ~&-y_1 ~& 0~ &0~& 0& \cdots& 0\cr
  -y_1~& x_2+y_1+y_2~ & -y_2~ & 0~ & 0 & \cdots &0\cr
0~ & -y_2~ &x_3+y_2+y_3~ & -y_3~ & 0 &\cdots & 0\cr
&&\cdots &&&&-y_{N-1}\cr
0&\cdots&0~&0~&0~&-y_{N-1}~&x_N+y_{N-1}\cr
\end{pmatrix}
\en
Let the matrix $O$ be an orthogonal matrix that diagonalizes $C$: $(OCO^T)_{\alpha\beta}
=\lambda_\alpha\delta_{\alpha\beta}$. The variable transformation
$\bm{Q}_\alpha\to O_{\alpha\beta}\lambda_\beta^{-1/2}\bm{Q}_\beta$ transforms the denominator into
\eq
D\to\sum_\alpha\bm{Q}_\alpha\bm{Q}_\alpha+\Delta,
\quad\quad
\Delta=(y_1+\cdots+y_{N-1})M^2-(x_1+\cdots+x_N)q_0^2\, .
\en
The multi-loop integral is then rewritten in the following form:
\eq
&&M_{4,N}^{\sf div}(q_0)
=\int_0^1\mathscr{D}_{xy}\int \frac{d^{Nd}\bm{Q}}{(2\pi)^{Nd}}\,
\dfrac{\det(C)^{-d/2}h_Nc_d^{(4)}}{D^{2N-1}}\,
  \left(N_4-\dfrac{6}{d+4}\,N_2+\dfrac{3}{(d+2)(d+4)}\,N_0\right)\, .
\nonumber\\
  &&
  \en
In order to simplify notations, we further introduce
the multiindex $A=(\alpha,i)$ and $B=(\beta,j)$.
Here $\alpha,\beta=1,\ldots N$
and $i,j=1,2,3$ label the loop momentum and the space index of the $Nd$-dimensional
vector $\bm{Q}=(\bm{Q}_1,\cdots,\bm{Q}_N)$. We further define
$v^{(1)}_\alpha=O_{1\alpha}\lambda_{\alpha}^{-1/2}$ and 
$v^{(N)}_\beta=O_{1\beta}\lambda_{\beta}^{-1/2}$. The terms in the
numerator of the Feynman integral can be written in the following compact notation
($K=0,2,4$)
\eq\label{eq:N_K}
N_K=v^{(1)}_{\alpha_1}\cdots v^{(1)}_{\alpha_4}\,v^{(N)}_{\beta_1}\cdots
v^{(N)}_{\beta_4}\,\Phi^{(K)}_{i_1\cdots i_4,j_1\cdots j_4}\, 
Q_{A_1}\cdots Q_{A_4}Q_{B_1}\cdots Q_{B_4}\, ,
\en
where
\eq
\Phi^{(0)}_{i_1\cdots i_4,j_1\cdots j_4}&=&\delta_{i_1i_2}\delta_{i_3i_4}\delta_{j_1j_2}\delta_{j_3j_4}\, ,
\nonumber\\[2mm]
\Phi^{(2)}_{i_1\cdots i_4,j_1\cdots j_4}&=&\delta_{i_1i_2}\delta_{j_1j_2}\delta_{i_3j_3}\delta_{i_4j_4}\, ,
\nonumber\\[2mm]
\Phi^{(4)}_{i_1\cdots i_4,j_1\cdots j_4}&=&\delta_{i_1j_1}\delta_{i_2j_2}\delta_{i_3j_3}\delta_{i_4j_4}\, .
\en
Since the denominator of the Feynman integral depends only on the $Nd$-dimensional vector squared, it is possible to average over the directions. We define
\eq\label{eq:Pi}
\langle Q_AQ_B\rangle&=&a_1\,Q^2\,\Pi^{(1)}_{AB}\, ,
\nonumber\\[2mm]
\langle Q_{A_1}Q_{A_2}Q_{B_1}Q_{B_2}\rangle&=&a_2\,Q^4\,\Pi^{(2)}_{A_1A_2B_1B_2}\, ,
\nonumber\\[2mm]
\langle Q_{A_1}Q_{A_2}Q_{A_3}Q_{B_1}Q_{B_2}Q_{B_3}\rangle&=&a_3\,Q^6\,\Pi^{(3)}_{A_1A_2A_3B_1B_2B_3}\, ,
\nonumber\\[2mm]
\langle Q_{A_1}Q_{A_2}Q_{A_3}Q_{A_4}Q_{B_1}Q_{B_2}Q_{B_3}Q_{B_4}\rangle&=&a_4\,Q^8\,\Pi^{(4)}_{A_1A_2A_3A_4B_1B_2B_3B_4}\, .
\en
It is clear that the quantities $\Pi$ are symmetric with respect to the permutations of
any of two indices. Furthermore,
\eq
\Pi^{(1)}_{AB}&=&\delta_{AB}\, ,
\nonumber\\[2mm]
\Pi^{(2)}_{A_1A_2B_1B_2}&=&\delta_{A_1A_2}\Pi^{(1)}_{B_1B_2}+\delta_{A_1B_1}\Pi^{(1)}_{A_2B_2}+\delta_{A_1B_2}\Pi^{(1)}_{A_2B_1}\, ,
\nonumber\\[2mm]
\Pi^{(3)}_{A_1A_2A_3B_1B_2B_3}&=&\delta_{A_1A_2}\Pi^{(2)}_{A_3B_1B_2B_3}+
\delta_{A_1A_3}\Pi^{(2)}_{A_2B_1B_2B_3}+\delta_{A_1B_1}\Pi^{(2)}_{A_2A_3B_2B_3}
\nonumber\\[2mm]
&+&
\delta_{A_1B_2}\Pi^{(2)}_{A_2A_3B_1B_3}+\delta_{A_1B_3}\Pi^{(2)}_{A_2A_3B_1B_2}\, ,
\nonumber\\[2mm]
\Pi^{(4)}_{A_1A_2A_3A_4B_1B_2B_3B_4}&=&
\delta_{A_1A_2}\Pi^{(3)}_{A_3A_4B_1B_2B_3B_4}
+\delta_{A_1A_3}\Pi^{(3)}_{A_2A_4B_1B_2B_3B_4}
\nonumber\\[2mm]
&+&\delta_{A_1A_4}\Pi^{(3)}_{A_2A_3B_1B_2B_3B_4}
+\delta_{A_1B_1}\Pi^{(3)}_{A_2A_3A_4B_2B_3B_4}
\nonumber\\[2mm]
&+&\delta_{A_1B_2}\Pi^{(3)}_{A_2A_3A_4B_1B_3B_4}
+\delta_{A_1B_3}\Pi^{(3)}_{A_2A_3A_4B_1B_2B_4}
\nonumber\\[2mm]
&+&\delta_{A_1B_4}\Pi^{(3)}_{A_2A_3A_4B_1B_2B_3}\, ,
\en
and so on. The normalization constants $a_1,\ldots,a_4$ can be easily determined, taking
the trace on both sides of Eq.~(\ref{eq:Pi})
\eq
a_1&=&\frac{1}{dN}\, ,\quad\quad
a_2=\frac{1}{dN(dN+2)}\, ,\quad\quad
a_3=\frac{1}{dN(dN+2)(dN+4)}\, ,
\nonumber\\[2mm]
a_4&=&\frac{1}{dN(dN+2)(dN+4)(dN+6)}\, .
\en
It remains to substitute the above formulae into Eq.~(\ref{eq:N_K}) and evaluate the numerator after averaging over the directions. After the lengthy but straightforward calculations
we arrive at a compact expression
\eq
M_{4,N}^{\sf div}(q_0)
=\int_0^1\mathscr{D}_{xy}\int \frac{d^{Nd}\bm{Q}}{(2\pi)^{Nd}}\,
\dfrac{\det(C)^{-d/2}h_Nc_d^{(4)}a_4\,(d-1)d(d+1)(d+6)\,f_{1N}^4\,Q^8}
{(Q^2+\Delta)^{2N-1}}\, ,\,\,\,
\en
where $f_{\alpha\beta}$ denotes the matrix element of $C^{-1}$: 
\eq
f_{\alpha\beta}=\sum_\gamma O_{\alpha\gamma}\lambda_\gamma^{-1}O_{\beta\gamma}=\det(C)^{-1}\mbox{Adj}(C)_{\alpha\beta}\, ,
\en
where $\mbox{Adj}(C)$ is the adjugate of the matrix $C$.
Carrying out the integration over $\bm{Q}$, we finally get
\eq
M_{4,N}^{\sf div}(q_0)
&=&\frac{g^{N-1}c_d^{(4)}a_4}{(4\pi)^{N(d/2-1)}}\,\frac{\Gamma(Nd/2+4)\Gamma(N(2-d/2)-5)}{\Gamma(Nd/2)}
\nonumber\\[2mm]
&\times&\int_0^1\mathscr{D}_{xy}\,\det(C)^{-d/2-4}\left(\mbox{Adj}(C)_{1N}\right)^4\Delta^{N(d/2-2)+5}\, .
\en
In order to carry out integrations, it is convenient to group the variables $x_i$ and $y_i$
separately:
\eq
x_1&=&\rho\tau_1,\quad\quad
x_2=\rho(1-\tau_1)\tau_2,\quad\cdots\quad
x_N=\rho(1-\tau_1)\cdots(1-\tau_{N-1})\, ,
\nonumber\\[2mm]
y_1&=&(1-\rho) t_1,\quad\quad
y_2=(1-\rho) (1-t_1)t_2,\quad\cdots\quad
y_{N-1}=(1-\rho) (1-t_1)\cdots(1-t_{N-2})\, ,
\nonumber\\[2mm]
s&=&(1-\rho)/\rho\, .
\en
As already mentioned, the quantity $M_{4,N}^{\sf div}(q_0)$ is ultraviolet divergent. It contains a simple pole in $(d-3)$ to all orders that can emerge in different places. For {\em even} values of $N$, the pole emerges in $\Gamma(N(2-d/2)-5)$, whereas the integration over the variable $s$ yields a finite result in dimensional regularization,
because the index of divergence is half-integer in $d=3$ dimensions. For {\em odd}
values of $N$, the prefactor is finite, and the pole emerges from the integration over $s$.
Integrating over the variables $\tau_i$ and $t_i$ does not produce any divergences.
All diagrams with $N>10$ are ultraviolet-finite. 

Separating the poles in multi-loop diagrams is a rather involved enterprise. On the other hand, applying the subtraction at threshold (or any other point below threshold, with a subtraction scale of order of $M$) can be done straightforwardly. We shall fix our renormalization
prescription by subtracting from the multi-loop diagram the
first few terms of the Taylor expansion in the variable $z=q_0^2/M^2$.
The polynomial of the fourth order in $z$ should be subtracted for $N=2$,
the polynomial of the third order for $N=3,4$, the polynomial of the second order
for $N=5,6$, and so on. It is clear that this procedure does not introduce large mass scale
and, hence, the coefficients of the modified effective range expansion in
$K_\ell^M(q_0^2)$ will be of natural size.

\subsection{Results of numerical calculations for $M_\ell(q_0)$}

In this section we shall present the results of numerical calculations for $\ell=0,4$. After
subtracting the ultraviolet divergences, the calculation over Feynman parameters
in $d=3$ dimensions were carried by using the VEGAS routine~\cite{Lepage:1977sw,Lepage:2020tgj}. The Lippmann-Schwinger equation~(\ref{eq:div-fin}) that determines $T_L^{\sf fin}$
is discretized on Gaussian mesh points and solved by using matrix inversion.

\begin{figure}[t]
  \begin{center}
    \includegraphics[width=0.35\paperwidth]{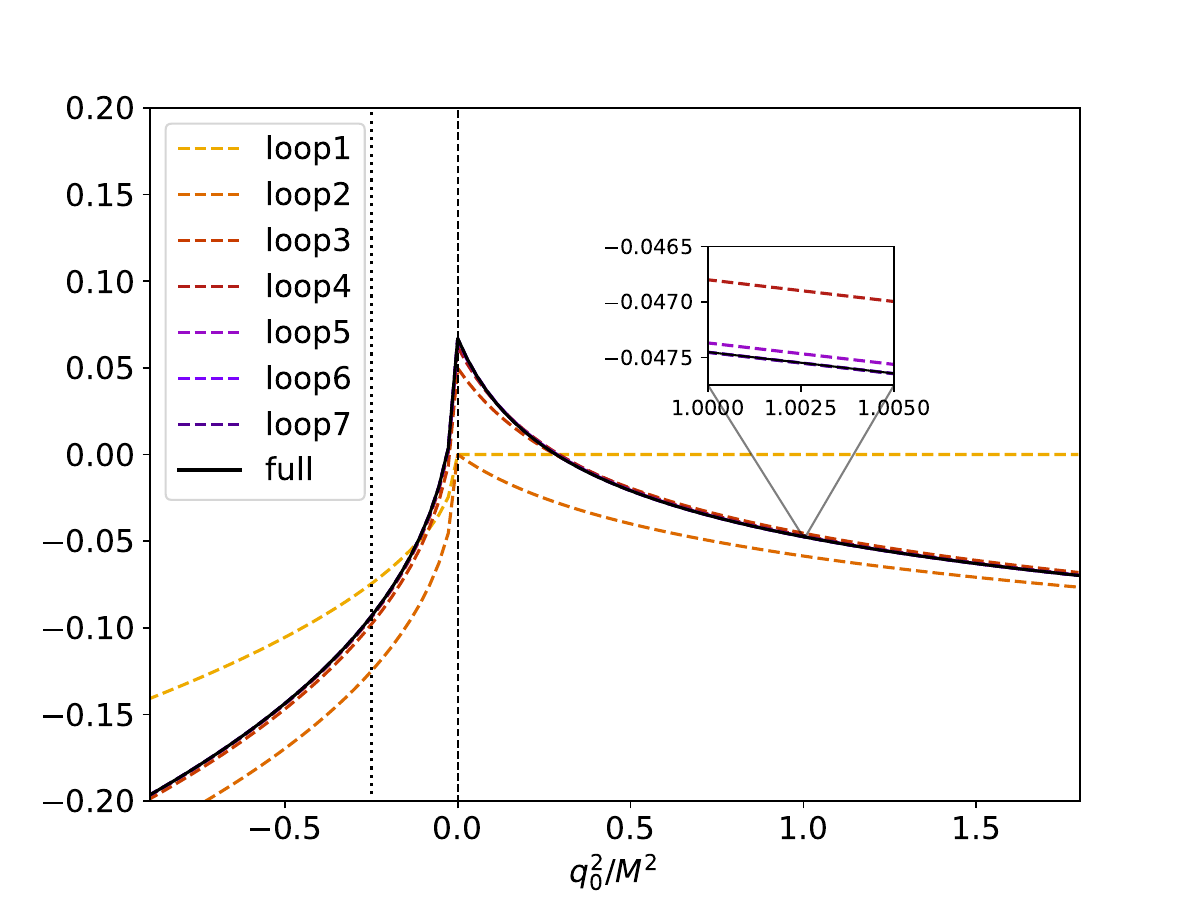}
    \includegraphics[width=0.35\paperwidth]{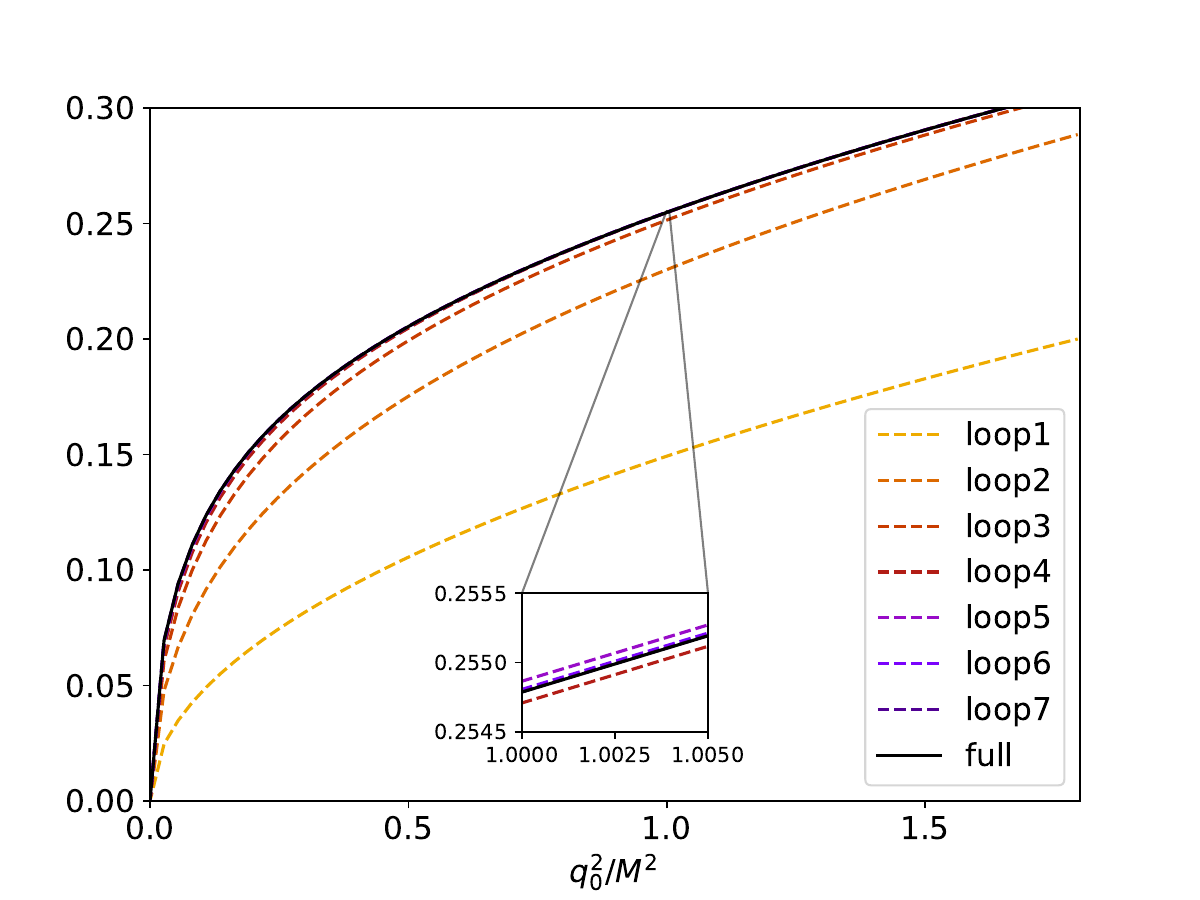}
    \caption{The real and imaginary parts of the
      function $M_\ell(q_0)$ for $\ell=0$. The imaginary part is zero below threshold.
      The full solution corresponds to the sum of the $M_\ell^{\sf div}(q_0)$ (up to two loops)
      and $M_\ell^{\sf fin}(q_0)$, see Eq.~(\ref{eq:div-fin}).
      For comparison, the perturbative result up to 7 loops is shown. The vertical lines in the left panel correspond to the elastic threshold and the beginning of the left-hand cut.
          }
    \label{fig:M0}
    \end{center}
    \end{figure}

In Fig.~\ref{fig:M0} the real and imaginary parts of the
function $M_\ell(q_0)$ for $\ell=0$ are plotted. The full solution $M_\ell(q_0)=M_\ell^{\sf div}(q_0)
+M_\ell^{\sf fin}(q_0)$ is given by a sum of two terms, where the former contains perturbative
contributions up to two loops. For comparison, we also plot the result of perturbative calculations
up to and including 7 loops. It is seen that, for a given value of the coupling $g$, the perturbative
series converges rapidly, so the result in 4 loops and higher becomes visually indistinguishable
from the full solution.

\begin{figure}[t]
  \begin{center}
    \includegraphics[width=0.35\paperwidth]{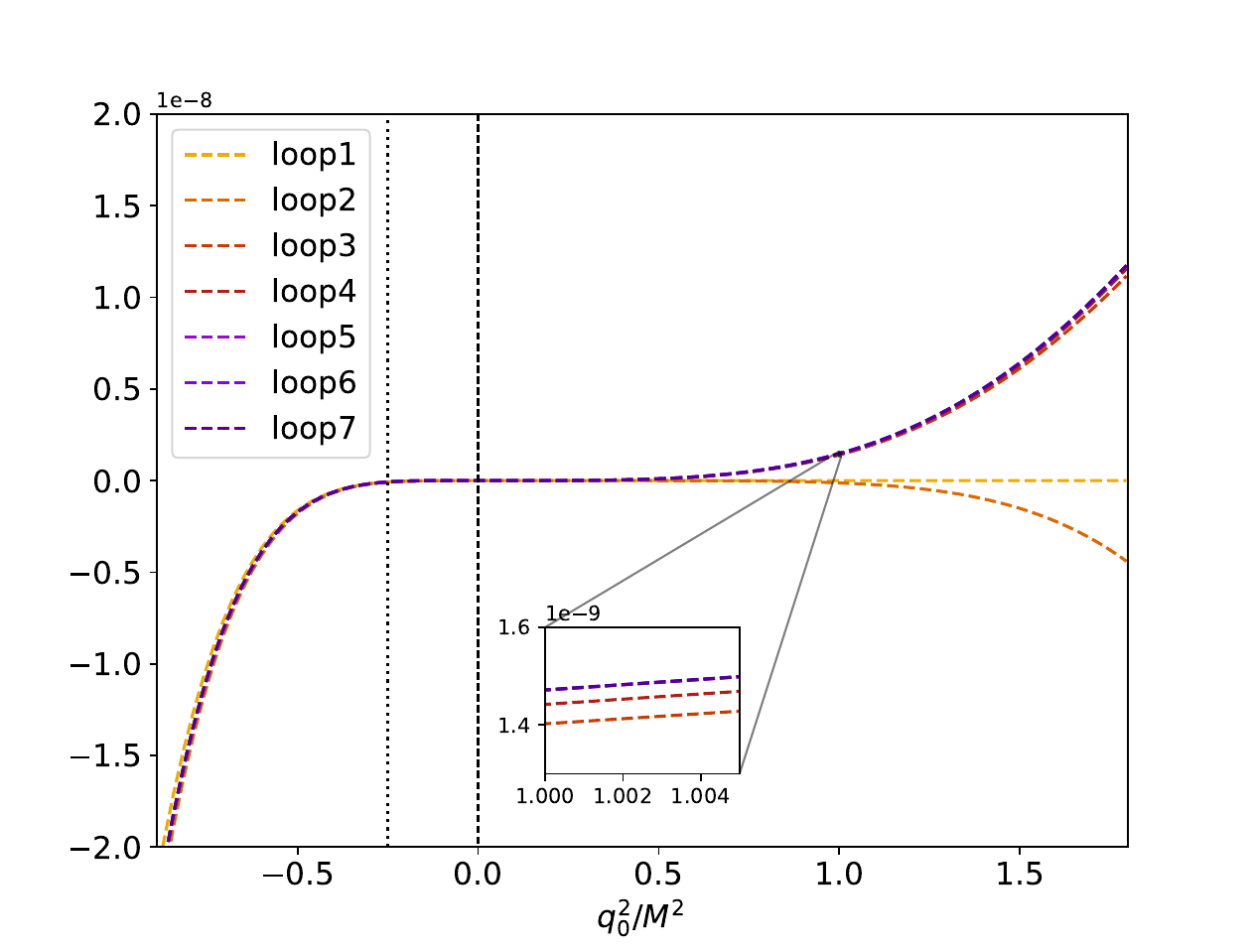}
    \includegraphics[width=0.35\paperwidth]{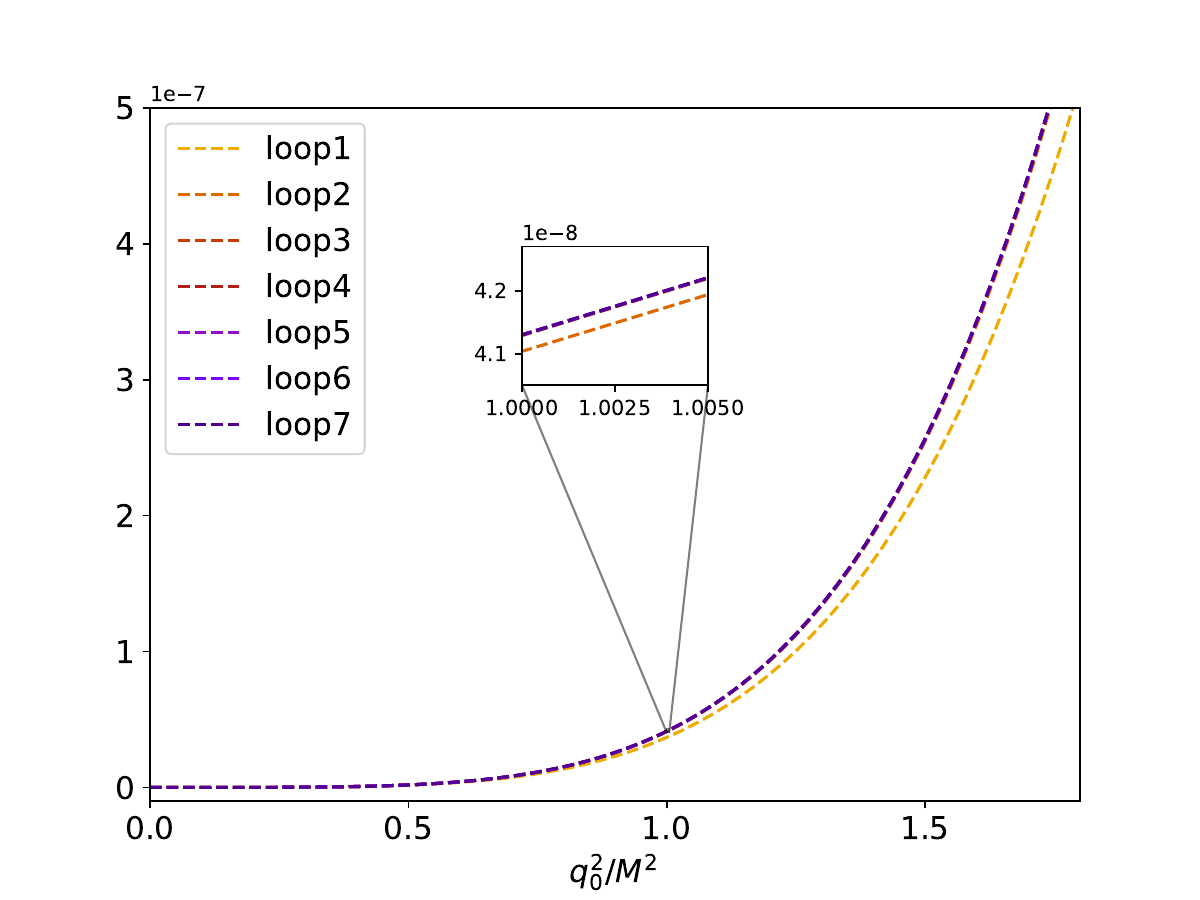}
    \caption{The real and imaginary parts of the
      function $M_\ell(q_0)$ for $\ell=4$. The imaginary part is zero below threshold.
      The perturbative expansion converges very rapidly,
      so that the contribution of
      $M_\ell^{\sf fin}(q_0)$ is invisible by the bare eye and is not shown (see also the
      zoo\-med-in windows). The vertical lines in the left panel correspond to the elastic threshold and the beginning of the left-hand cut.
    }
    \label{fig:M4}
    \end{center}
    \end{figure}

In Fig.~\ref{fig:M4} the result of the calculation in case $\ell=4$ is shown. Now,
$M_\ell^{\sf div}(q_0)$ contains 10 terms, and $M_\ell^{\sf fin}(q_0)$ is so small that it cannot
be distinguished with a bare eye. As one sees from the figure, the convergence is again very good and, after calculating a few loops, the full solution does not change anymore. Note that oscillations in the real part are an artifact of the subtraction and do not have an effect on the convergence.

Last but not least, note that only repulsive interactions are considered in the present
paper. In case of attractive interactions, bound states can emerge in the spectrum and
the Born series is no longer convergent. In our approach, these bound state poles show up in
$M_\ell^{\sf fin}(q_0)$.

\subsection{Coulomb interactions}

In the limit $M\to 0$, our long-range force reduces to the Coulomb force, for which
the analytic solution is known~\cite{Bethe:1949yr,Chew:1949zz}. In this case, we have
to identify the dimensionful constant $g$ with $g=2\mu_r\alpha$, where $\mu_r$ is the
reduced mass and $\alpha$ denotes the fine structure constant. The quantity $M_\ell(q_0)$
becomes the function of the dimensionless variable $\eta=\alpha\mu_r/q_0$. Restricting
ourselves, for simplicity, to the case $\ell=0$, we get
\eq\label{eq:BCG}
M_0(q_0)=2\alpha\mu_r\biggl\{\eta^2\sum_{n=1}^\infty\frac{1}{n(n^2+\eta^2)}-\log\eta+\Gamma'(1)\biggr\}\, .
\en
We have checked that, in the limit $M\to 0$, the expression in the curly brackets
in Eq.~(\ref{eq:ell=0}) indeed yields the term
$-\log\eta$ up to an inessential constant contribution. Fixing of this contribution amounts to
setting the renormalization prescription. Higher order terms in the expansion in $\eta$
are obtained from the multi-loop integrals. The loops with odd $N$ are purely imaginary
and do not contribute. We have (numerically) checked that the first few coefficients 
of the expansion of $M_\ell(q_0)$ indeed reproduce those obtained from Eq.~(\ref{eq:BCG}) with a very good precision. Thus, our expression for $M_\ell(q_0)$ indeed reduces
to the known one from Refs.~\cite{Bethe:1949yr,Chew:1949zz} in the massless limit.

\subsection{Modified effective range expansion function}

As mentioned above, the modified quantization condition allows one to extract
the modified effective range function $K_\ell^M(q_0^2)$ both above and below threshold
and, in particular, in the left-land cut region where the standard L\"uscher approach fails.
Here we wish to construct $K_\ell^M(q_0^2)$ in the whole energy range and ensure that
it is smooth and real everywhere, in contrast to the standard $K$-matrix. We carry
out calculations in the model described by the potential (\ref{eq:potential_toy})
and restrict ourselves, for simplicity, to the case $\ell=0$.

It should be pointed out that the analytic continuation of the solution of the
Lippmann-Schwinger equation below threshold is by no means a trivial affair. As
discussed, e.g., in Ref.~\cite{Oller:2018zts}, when $q_0^2<-M^2$,
the path of momentum integration in the Lippmann-Schwinger integration, which originally runs from $0$ to infinity, should be deformed in order to avoid the singularities of the potential. The left-hand cut starts higher, at $q_0^2=-M^2/4$. Since the analysis in the left-hand cut region is our primary interest in this paper, we restrict the energy interval by
$q_0^2>-M^2$. Albeit, with some additional effort, the analysis can be extended to
the larger region $q_0^2>-M_S^2/4$. 

\begin{figure}[t]
  \begin{center}
    \includegraphics*[width=8.cm]{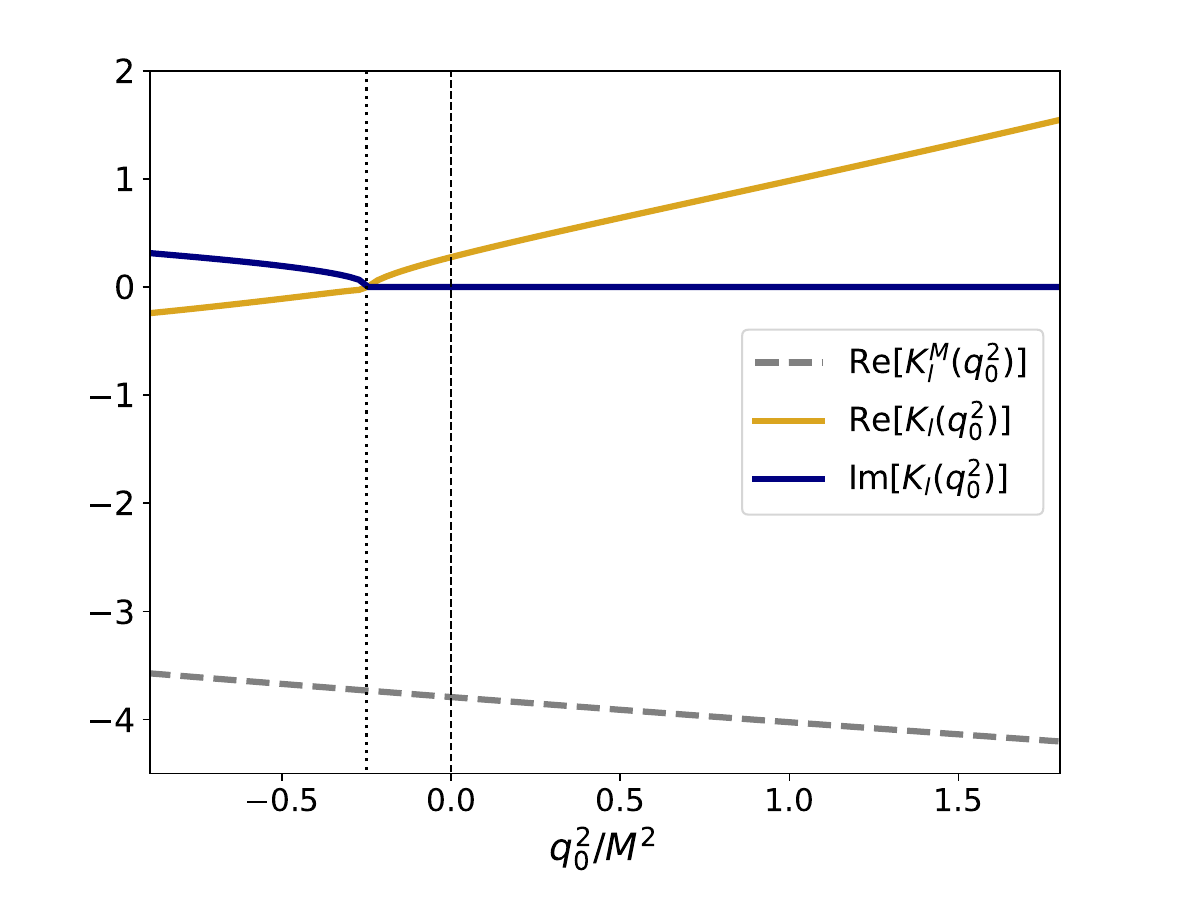}
    \caption{Real and imaginary parts (in arbitrary units)
      of the standard effective range function
      $K_\ell(q_0^2)$ (solid lines) vs. the real part of the modified effective range
      function $K_\ell^M(q_0^2)$ (dashed line) in the S-wave ($\ell=0$).
      The imaginary part of $K_\ell^M(q_0^2)$ is
    zero everywhere in the interval considered. The position of the left-hand threshold is shown by a vertical dotted line.}
    \label{fig:K_ell-M}
  \end{center}
  \end{figure}

Another subtle issue is the definition of the effective range expansion below threshold.
The definition (\ref{eq:modified}) applies above threshold. In the interval
$-M^2<q_0^2<0$ we use the following definition:
\eq
K_\ell^M(q_0^2)=M_\ell(q_0)+\frac{4\pi q_0^{2\ell}}{|f_\ell(q_0)|^2}\,\frac{1}{T_\ell(q_0)-T^L_\ell(q_0)}\, ,
\en
where $T_\ell(q_0),T^L_\ell(q_0)$ denote the full and the long range fully on-shell
scattering amplitudes
(these amplitudes become complex in the left-hand cut region, however the difference
$T_\ell(q_0)-T^L_\ell(q_0)$ stays real). The Jost function is also real in the considered
subthreshold interval and is given by
\eq
f_\ell(q_0)^{-1}=1+\frac{1}{q_0^\ell}\,\int_0^\infty\frac{p^2dp}{2\pi^2}\,
\frac{p^\ell}{p^2-q_0^2}\,T^L_\ell(p,q_0;q_0^2)\, ,\quad\quad
T^L_\ell(q_0)=T^L_\ell(p,q_0;q_0^2)\, .
\en
In Fig.~\ref{fig:K_ell-M} we show the modified effective range function $K_\ell^M(q_0^2)$
vs. the standard one $K_\ell(q_0^2)$. The difference is clearly visible. While the standard
function displays a singular structure at the left-land threshold and becomes complex
below it, the modified effective range function is almost linear in the whole interval
considered. Moreover, it does not demonstrate any sign of a singular behavior even
very close to $q_0^2=-M^2$, in agreement with the claim that the convergence radius of
the modified effective range expansion is set by the short-range scale $M_S$.

\section{Conclusions}

\begin{itemize}

\item[i)] In the present paper we discussed the numerical implementation of the modified
  L\"uscher equation, proposed  in Ref.~\cite{Bubna:2024izx}. We chose
  to demonstrate the essential features of the implementation, using a simple toy model:
  two spinless non-relativistic particles interacting through a sum of two Yukawa potentials,
  having different ranges. Furthermore, the center-of-mass frame was chosen from the
  beginning, and we restricted ourselves to the case of the $A_1^+$ irrep of the octahedral
  group only. All these are purely technical restrictions and can be easily removed in the
  analysis of real lattice data.

\item[ii)] In addition to the above, we assumed that the long-range force is perturbative
  and does not create bound states/low-lying resonances alone. In contrast to the
  other assumptions, this one is more restrictive and might require additional scrutiny.
  For the time being, we however stick to this assumption, because it is justified for the real
  physical systems which are studied on the lattice at present.

\item[iii)] We went step by step and presented a simple but rather accurate and fast
  algorithm for the calculation of the modified L\"uscher zeta-function  shown in
  Fig.~\ref{fig:loops}. This quantity is ultraviolet-divergent as its renormalization is a
  non-trivial issue, especially in higher partial waves. From the point of numerical
  algorithms, a cutoff is a preferred choice. However, cutoff regularization leads to the very
  large subtraction terms in higher partial waves that affects the accuracy of calculations.
  From this point of view, dimensional regularization is preferred because in this
  case the subtraction terms are of natural size in all partial waves. We
  derive explicit expressions for the renormalized $n$-loop exchange diagrams in dimensional regularization, using a subtraction scheme that does not produce unnaturally
  large polynomial contributions.

\item[iv)] The solutions of the quantization condition (both modified and standard ones)
  are compared with the exact finite-volume spectrum of the Hamiltonian, calculated
  in the plane-wave basis. The results confirm all our expectations. First of all, exponential
  corrections are very small for all levels except the ground state, where they are increased
  owing to the proximity of the left-hand cut. Furthermore, the convergence of the
  partial-wave expansion is very good in the modified case and poor in the standard case.
  Indeed, the standard case implies the partial-wave expansion of the whole potential, whereas
  in the modified case only the short-range potential is expanded in partial waves, the long-range
  part is treated in the plane-wave basis as a whole.

\item[v)] If the modified L\"uscher zeta-function for a known long-range potential is calculated and tabulated in advance, the analysis of lattice data proceeds exactly in the same manner as in the case of the standard L\"uscher equation. The truncation in partial waves, which involves short-range potential only, is justified.
This potentially renders the proposed approach very convenient for the analysis of lattice data.

\end{itemize}

{\em Acknowledgments:}
The authors especially acknowledge the contribution from Fabian M\"uller,
who intensively collaborated with us at the early stage of the project. We thank E. Epelbaum, A. Gasparyan, B. Metsch and Yan Li for useful discussions.
	The work of R.B, F.M. and A.R. was  funded in part by
	the Deutsche Forschungsgemeinschaft
	(DFG, German Research Foundation) – Project-ID 196253076 – TRR 110 and by the Ministry of Culture and Science of North Rhine-Westphalia through the NRW-FAIR project.
	A.R., in addition, thanks the Chinese Academy of Sciences (CAS) President's
	International Fellowship Initiative (PIFI) (grant no. 2024VMB0001) for the
	partial financial support.
	The work of J.-Y.P. and J.-J.W. was supported by the National Natural Science Foundation of China (NSFC) under Grants No. 12135011, 12175239, 12221005, and by the National Key R\&D Program of China under Contract No. 2020YFA0406400, and by the Chinese Academy of Sciences under Grant No. YSBR-101.
    H.-W.H. was supported by Deutsche Forschungsgemeinschaft (DFG, German Research Foundation) under Project ID 279384907 – SFB 1245 and by the German Federal Ministry of Education and Research (BMBF) (Grant No. 05P24RDB).
    B.-L. H.  was supported by the DFG through the Research Unit  ``Photon--photon interactions in the Standard Model and beyond'' (Projektnummer 458854507 -- FOR 5327).

\appendix
\section{Partial-wave contributions to the energy shift of a given state}
\label{app:PW}
In this appendix, we give a qualitative explanation for the uneven convergence of the partial-wave expansion, which is observed in Fig.~\ref{fig:HEFT}.
We shall namely argue that the pattern of the convergence in different excited states can be understood solely on the basis of a symmetry argument. 

In the beginning, let us note that the energy shifts from the unperturbed states are generally small, so first-order perturbation theory captures the bulk of the effect. These unperturbed states represent the
finite-volume momentum eigenstates $|\bm{n}\rangle$ that satisfy the completeness condition
\begin{align}
\sum_{\bm{n}\in\mathbb{Z}^{3}}|\bm{n}\rangle\langle\bm{n}| & =\mathbbm{1}.
\end{align}
Let us now introduce the {\em shells} -- sets of vectors generated from a given vector 
$|\bm{n}\rangle$ by the action of all elements of the octahedral group $O_h$.
 We label these shells by index $r$. It is clear that the matrix representations of $O_h$ can be realized in the linear space spanned by all vectors belonging to a given shell. These representations are in general reducible, and can be decomposed into the different irreps.
There are seven fundamental types of shells 
(see, e.g.~\cite{Doring:2018xxx,Meng:2021uhz,Li:2019qvh}), classified according to how many components of the three-momentum share the same magnitude. The decomposition of the representations of the group $O_h$ for these shells are given by
\eq
(0,0,0)&:&\quad A_{1}^{+}\,,\nonumber\\[2mm]
(0,0,c) &:&\quad A_{1}^{+}\oplus E^{+}\oplus T_{1}^{-}\, ,\nonumber\\[2mm]
(0,b,b) &:&\quad  A_{1}^{+}\oplus E^{+}\oplus T_{1}^{-}\oplus T_{2}^{+}\oplus T_{2}^{-}\,,\nonumber\\[2mm]
(0,b,c)&:&\quad A_{1}^{+}\oplus A_{2}^{+}\oplus 2E^{+} 
\oplus T_{1}^{+}\oplus 2T_{1}^{-}\oplus T_{2}^{+}\oplus 2T_{2}^{-}\, ,\nonumber\\[2mm]
(a,a,a)&:&\quad A_{1}^{+}\oplus A_{2}^{-}\oplus T_{1}^{-} 
\oplus T_{2}^{+}\, ,\nonumber\\[2mm]
(a,a,c)&:&\quad A_{1}^{+}\oplus A_{2}^{-}\oplus E^{+}\oplus
E^{-}\oplus T_{1}^{+}\oplus2T_{1}^{-}\oplus2T_{2}^{+}
\oplus T_{2}^{-}\,,\nonumber\\[2mm]
(a,b,c) &:&\quad A_{1}^{+}\oplus A_{1}^{-}\oplus A_{2}^{+}\oplus A_{2}^{-}\oplus2E^{+}\oplus2E^{-}
\oplus3T_{1}^{+}\oplus3T_{1}^{-}\oplus3T_{2}^{+}\oplus 3T_{2}^{-}\,.
\en
We denote the basis vectors for each irrep 
by $|\Gamma t,\alpha;r\rangle$. Here $\Gamma$
labels the irrep of $O_{h}$, $\alpha$ distinguishes the basis vectors
within that irrep, $r$ labels shells and $t$
enumerates the multiplicity of the irrep $\Gamma$~\cite{Li:2019qvh}.

Let us now evaluate the finite-volume energy shift of an unperturbed state $|\bm{n}\rangle$. When the interaction is switched on, the level splitting occurs, and the states belonging to the different irreps are shifted differently. For a given interaction potential $V$, the leading-order perturbative energy shift of a state transforming in the irrep $\Gamma$ is obtained by diagonalization of the following matrix in the space of indices $t,t'$:
\begin{align}\label{eq:Vtt1}
V_{tt'}^{(\Gamma)}(r) & =\langle\Gamma t,\alpha;r|V|\Gamma t',\alpha;r\rangle.
\end{align}
Here $\alpha$ is arbitrary due to the Wigner-Eckart theorem. 

Furthermore, the interaction
potential can be expanded in partial waves: 
\begin{align}
\langle\bm{n}_1|V|\bm{n}_2\rangle & =4\pi\sum_{\ell m}Y_{\ell m}(\hat{\bm{n}}_1)V_{\ell}(p_1,p_2)Y_{\ell m}^{*}(\hat{\bm{n}}_2),
\end{align}
where $p_i=\dfrac{2\pi}{L}|\bm{n}_i|$.
Therefore, the matrix element in Eq.~(\ref{eq:Vtt1}) is given by
\eq
V_{tt'}^{(\Gamma)}(r) & =&\sum_{\bm{n}_{1},\bm{n}_{2}\in\text{shell-}r}\langle\Gamma t,\alpha;r|\bm{n}_{1}\rangle\langle\bm{n}_{1}|V|\bm{n}_{2}\rangle\langle\bm{n}_{2}|\Gamma t',\alpha;r\rangle\nonumber \\[2mm]
 & =&4\pi\sum_{\ell m}V_{\ell}(p_1,p_2)\sum_{\bm{n}_{1},\bm{n}_{2}\in\text{shell-}r}\langle\Gamma t,\alpha;r|\bm{n}_{1}\rangle Y_{\ell m}(\hat{\bm{n}}_{1})Y_{\ell m}^{*}(\hat{\bm{n}}_{2})\langle\bm{n}_{2}|\Gamma t',\alpha;r\rangle\,.
\en
Note that since $\bm{n}_1$ and $\bm{n}_2$ both belong to the 
shell-$r$, $p_1=p_2$ is the magnitude of the three-momentum in the shell-$r$. 

Next, we introduce the finite-volume partial-wave states 
$|\ell,m;r\rangle$, defined through projecting the plane-wave basis onto the spherical harmonics~\cite{Li:2019qvh}
\begin{align}
|\ell,m;r\rangle & =\sum_{\bm{n}\in\text{shell-}r}\sqrt{4\pi}\,Y_{\ell m}(\hat{\bm{n}})\,|\bm{n}\rangle\, ,
\end{align}
and define 
\begin{align}
|\Gamma t\ell,\alpha;r\rangle & =\sum_{m}c_{\ell m}^{\Gamma \alpha t}|\ell,m;r\rangle\, ,
\end{align}
where $c_{\ell m}^{\Gamma \alpha t}$ denotes the pertinent Clebsh-Gordan coefficient, see, e.g., Eq.~(\ref{eq:A-Gamma}). Since the transformation between $|\ell,m;r\rangle$ and $|\Gamma t\ell,\alpha;r\rangle$
is unitary, we have 
\begin{align}
\sum_{m}|\ell,m;r\rangle\langle\ell,m;r| & =\sum_{\Gamma t\alpha}|\Gamma t\ell,\alpha;r\rangle\langle\Gamma t\ell,\alpha;r|\, .
\end{align}
This means that 
\eq
V_{tt'}^{(\Gamma)}(r) & =&4\pi\sum_{\ell m}V_{\ell}(p,p)\sum_{\bm{n}_{1},\bm{n}_{2}\in\text{shell-}r}\langle\Gamma t,\alpha;r|\bm{n}_{1}\rangle Y_{\ell m}(\hat{\bm{n}}_{1})Y_{\ell m}^{*}(\hat{\bm{n}}_{2})\langle\bm{n}_{2}|\Gamma t',\alpha;r\rangle\nonumber \\[2mm]
 & =&\sum_{\ell m}V_{\ell}(p,p)\langle\Gamma t,\alpha;r|\ell,m;r\rangle\langle\ell,m;r|\Gamma t',\alpha;r\rangle\nonumber \\[2mm]
 & =&\sum_{\ell}V_{\ell}(p,p)\sum_{\Gamma^{'}t''\beta}\langle\Gamma t,\alpha;r|\Gamma^{'} t''\ell,\beta;r\rangle\langle\Gamma^{'}t''\ell,\beta;r|\Gamma t',\alpha;r\rangle\nonumber \\[2mm]
 & =&\sum_{\ell}V_{\ell}(p,p)\sum_{t''}\langle\Gamma t,\alpha;r|\Gamma t''\ell,\alpha;r\rangle\langle\Gamma t''\ell,\alpha;r|\Gamma t',\alpha;r\rangle.
\en
Now let us define
\begin{align}
G_{\ell,tt'}^{(\Gamma)}(r) & =\sum_{t''}\langle\Gamma t,\alpha;r|\Gamma t''\ell,\alpha;r\rangle\langle\Gamma t''\ell,\alpha;r|\Gamma t',\alpha;r\rangle.
\end{align}
The quantity $G$ was calculated in Ref.~\cite{Li:2019qvh}.
In order to understand the result presented in Fig.~\ref{fig:HEFT},
we restrict ourselves to the irrep $\Gamma=A_{1}^{+}$ and truncate the partial-wave expansion, keeping only $\ell=0,4$. In this case, the multiplicities $t,t',t''$ and the label of the basis vector, $\alpha$,
all take a single value. For simplicity, we shall ignore these indices in the following. The leading‐order energy shift then reduces to
\eq
\Delta E^{(A_{1}^{+})}(r) =V^{(A_{1}^{+})}(r)
 =G_{0}^{(A_{1}^{+})}(r)V_0(p,p)+G_{4}^{(A_{1}^{+})}(r)V_4(p,p)+\cdots\, .
\en
The quantity $G$ is given by 
\begin{align}
G_{\ell}^{(A_{1}^{+})}(r) & =|\langle A_{1}^{+};r|A_{1}^{+}\ell;r\rangle|^{2}.
\end{align}
This quantity can be evaluated directly, resulting in
\begin{align}
\begin{array}{l l l}
(000) &:\quad G_{0}^{(A_{1}^{+})}=1\, ,\quad&G_{4}^{(A_{1}^{+})}=0\, ,\\[2mm]
(001) &:\quad G_{0}^{(A_{1}^{+})}=6\, ,\quad&G_{4}^{(A_{1}^{+})}=31.5\, ,\\[2mm]
(011) &:\quad G_{0}^{(A_{1}^{+})}=12\, ,\quad&G_{4}^{(A_{1}^{+})}=3.94\, ,\\[2mm]
(012) &:\quad G_{0}^{(A_{1}^{+})}=24\, ,\quad&G_{4}^{(A_{1}^{+})}=5.04\, ,\\[2mm]
(111) &:\quad G_{0}^{(A_{1}^{+})}=8\, ,\quad&G_{4}^{(A_{1}^{+})}=18.67\, ,\\[2mm]
(112) &:\quad G_{0}^{(A_{1}^{+})}=24\, ,\quad&G_{4}^{(A_{1}^{+})}=7.88\, ,\\[2mm]
(123) &:\quad G_{0}^{(A_{1}^{+})}=48\, ,\quad&G_{4}^{(A_{1}^{+})}=15.75\, .
\end{array}
\end{align}
The seven cases above correspond to the seven types of the momentum shells.
Relatively large $G_{4}^{(A_{1}^{+})}$ factors associated
with the shells $(001)$ and $(111)$ are responsible for sizable $G$-wave contributions
to the first and third excited states. And, {\it vice versa,} much smaller
$G_{4}^{(A_{1}^{+})}$ factors for $(011)$, $(012)$ and $(112)$ shells
explain, why the $G$-wave corrections are negligible in the second,
fifth, and sixth excited states, see Fig.~\ref{fig:HEFT}.

\bibliographystyle{unsrt}
\bibliography{refmod}

\end{document}